\documentclass[12pt]{article}

\usepackage{amsfonts}
\usepackage{bbm}

\newcommand{\bmat}{\left(\begin{array}}
\newcommand{\emat}{\end{array}\right)}
\newcommand{\uno}{\mathbbm{1}}

\def\ts{\textstyle}
\def\ds{\displaystyle}

\def\a{\alpha}
\def\b{\beta}

\def\Om{\Omega}

\def\-{\hphantom{-}}
\def\ov{\overline}
\def\s2{\frac{1}{\sqrt2}}

\def\oh{\frac{1}{2}}
\def\beq{\begin{equation}}
\def\eeq{\end{equation}}
\def\beqa{\begin{eqnarray}}
\def\eeqa{\end{eqnarray}}
\def\so{{\rm SO}}
\def\us{{\rm USp}}
\def\u{{\rm U}}
\def\su{{\rm SU}}
\def\tr{{\rm tr \,}}
\def\Tr{{\rm Tr \,}}

\def\T{{\rm T}}
\def\D{{\rm D}}
\def\Z{{\mathbb Z}}

\def\cc{{\mathcal C}}

\def\cg{{\mathcal G}}
\def\ck{{\mathcal K}}
\def\cam{{\mathcal M}}

\def\cz{{\mathcal Z}}

\def\op{{\Omega^\prime}}
\def\nm{{\ts{\frac{N}2}}}
\def\deq#1{\mbox{$D$=#1}}

\def\Dsl{\,\raise.15ex\hbox{/}\mkern-13.5mu D} %this one can be subscripted
\def\r#1{\mbox{{\bf #1}}}

\newcommand{\msm}[1]{\mbox{\footnotesize$#1$}}
\newcommand{\smim}[1]{\mbox{\footnotesize#1}}
\newcommand{\ket}[1]{| #1 \rangle}
\newcommand{\inpar}[1]{\left(#1\right)}
\newcommand{\inmod}[1]{\left|#1\right|}
\newcommand{\inll}[1]{\left\{#1\right\}}
\newcommand{\incor}[1]{\left[#1\right]}
\newcommand{\repf}{\square}
\newcommand{\repfc}{\overline{\square}}
\newcommand{\reps}{\square\!\square}
\newcommand{\repa}{\raisebox{4pt}{$\square$}\hspace*{-0.325cm}{\raisebox{-4pt}{$\square$}}}
\newcommand{\repsc}{\overline{\square\!\square}}
\newcommand{\repac}{\overline{\raisebox{4pt}{$\square$}\hspace*{-0.325cm}{\raisebox{-4pt}{$\square$}}}}

\begin{document}
\pagestyle{empty}
\begin{flushright}
{\tt IFT-UAM/CSIC-06-26}
\end{flushright}
\vspace*{2cm}

\vspace{0.3cm}
\begin{center}
{\LARGE \bf A class of non-supersymmetric orientifolds}\\[1cm]
Anamar\'{\i}a Font
\footnote{On leave from Departamento de F\'{\i}sica, Facultad de Ciencias,
Universidad Central de Venezuela, A.P. 20513, Caracas 1020-A, Venezuela.}\\[0.2cm]
{\it  Instituto de F\'{\i}sica Te\'orica C-XVI,
Universidad Aut\'onoma de Madrid,\\[-1mm]
Cantoblanco, 28049 Madrid, Spain. }\\[0.3cm]
and \\[0.3cm]
Jos\'e Antonio L\'opez \\[0.2cm]
{\it  Centro de F\'{\i}sica Te\'orica y Computacional, Facultad de Ciencias, \\[-1mm]
Universidad Central de Venezuela,\\[-1mm]
A.P. 47270, Caracas 1041-A, Venezuela. }\\[1cm]
\normalsize{\bf Abstract} \\[3mm]
\end{center}

\begin{center}
\begin{minipage}[h]{15.5cm}
\normalsize{
We study type IIB orientifolds on ${\rm T}^{2d}/\Z_N$ with supersymmetry broken by the
compactification. We determine tadpole cancellation conditions including antibranes and
considering different actions for the parity $\Omega$. Using these conditions we then
obtain the spectrum of tachyons and massless states. Various examples with $N$ even
correspond to type 0B orientifolds.
}
\end{minipage}
\end{center}

\newpage

%----------------------------------------------------------------------%
%  Resetting of counters
%----------------------------------------------------------------------%
\setcounter{page}{1} \pagestyle{plain}
\renewcommand{\thefootnote}{\arabic{footnote}}
\setcounter{footnote}{0}
%----------------------------------------------------------------------%
%  Paper begins
%----------------------------------------------------------------------

\section{Introduction}

Several ways to build non-supersymmetric open string models have been developed
in recent years \cite{nosusy}. One method is to start with orientifolds
of the ten-dimensional type 0 strings.  
The purpose of this note is to pursue this approach further. To this end we
will discuss a class of type IIB orientifolds on ${\rm T}^{2d}/\Z_N$ in which 
supersymmetry is completely broken by the $\Z_N$ generator $\theta$. 
The underlying idea is that when no compactification is
involved, and $\theta$ is a $\Z_2$ twist corresponding to $(-1)^{F_S}$, where
$F_S$ is space-time fermion number, one obtains the ten-dimensional
orientifolds of type 0B string that were first studied in \cite{bs, Sagnotti, bg}.
Some lower-dimensional cases can be regarded as compactifications of the
\deq10 type 0B orientifolds but the generic situation is just that of a
non-supersymmetric type IIB orientifold in $\deq(10-2d)$ dimensions.
Our motivation is to provide a unified description of these constructions.
We will consider both the standard world-sheet parity $\Omega$ and the modified
$\op$ defined as
\beq
\op= \Omega (-1)^{f_R}  \  ,
\label{omprime}
\eeq
where $f_R$ is the right-moving
world-sheet fermion number. In \deq10, $\op$ leads to a tachyon
free model \cite{Sagnotti} and this can also occur upon compactification
for some specific Abelian twists \cite{Angelantonj98, bfl, bk, kf, dm}.

As usual, the orientifold projection introduces
${\rm O}p$-planes whose RR charges have to be cancelled by
adding appropriate $\D p$-branes \cite{pol, as}. Since we
are breaking supersymmetry, we might as well add anti
$\D p$-branes ($\ov{\D p}$), just as in toroidal orientifolds
in which $\Z_N$ preserves supersymmetry \cite{au}. 
In general, there is a further sign freedom in the
M\"obius strip amplitude, or equivalently in the charge and tension
of the ${\rm O}p$-planes, that
allows either an orthogonal or a symplectic projection on the gauge Chan-Paton
factors \cite{as, dms}. For instance, in \deq10,
with $\Omega$ and no $(-1)^{F_S}$ twist,
and including $k$ $\ov{\D 9}$-branes, there can be type I-like
models with group $\so(32+k)\times \so(k)$, or
$\us(k\!-\!32)\times \us(k)$ Sugimoto models \cite{Sugimoto}.
Including the $(-1)^{F_S}$ twist gives instead the
$\incor{\so(k_1) \times \so(k_2)}^2$ models of \cite{bs, Sagnotti, bg},
with cancelled NSNS tadpoles if $k_1+k_2=32$, or
new $\incor{\us(k_1) \times \us(k_2)}^2$ non-supersymmetric orientifolds.

In this work we will explain the general construction of
$D \leq 10$ examples. In particular, we will derive tadpole
cancellation conditions valid in non-supersymmetric 
as well as in supersymmetric orientifolds studied in the past 
\cite{gp, gj, dp, Zwart, afiv, ah}. 
For lower dimensional $\D p$-branes our results carefully take into
account their location in the transverse directions. 
As remarked in \cite{abiu}, only such complete tadpole conditions
can guarantee absence of anomalies.

Since the $\Z_N$ action breaks supersymmetry explicitly, the resulting
gravity and gauge fields are purely bosonic. However, an interesting property
of these models is the appearance of massless charged fermions in the open string
sector. The spectrum typically contains tachyons from the closed twisted
sectors and also from open string sectors when $\overline{\D p}$-branes are present.
Moreover, there are generically uncancelled NSNS tadpoles although they could be
absent in some cases.
Nonetheless, we believe that this class of non-supersymmetric orientifolds
deserves further investigation. Tachyons and NSNS tadpoles are a common
feature of many non-supersymmetric orientifolds but it is conceivable that
a consistent theory could emerge after some stabilization process.

This note is organized as follows.
In section \ref{sgen} we review the building blocks of our non-supersymmetric
orientifolds. In the next sections we analyze the
various orientifolds classified according to the ${\rm O}p$-planes present.
In each case we first obtain tadpole cancellation conditions and then
provide the spectrum of tachyonic and massless states in several examples.
The 1-loop amplitudes needed to extract the RR tadpoles are explained
in the appendix. Some final remarks are stated in section \ref{sfin}.

\section{Generalities}
\label{sgen}

In this section we go over the basic concepts and notation needed to
describe our class of models.
We consider orientifolds with quotient group $\cg=(\uno+\Omega_P)\Z_N$,
where $\Omega_P$ is either the standard $\Omega$ or $\Omega^\prime$.
More details about the $\Z_N$ action will be presented in section \ref{sszn}.

Taking the quotient by $\cg$ introduces non-orientable Riemann surfaces in the
perturbative expansion. The one-loop vacuum-to-vacuum amplitude on the Klein
bottle ($\ck$) has generic tadpole divergences created by orientifold planes
charged under RR potentials. The natural way to
cancel such tadpoles is to add $\D p$-branes of opposite
RR charges. Open strings with ends on branes have one-loop
amplitudes on the cylinder ($\cc_{pq}$) and the M\"obius strip ($\cam_p$)
whose divergences cancel that of $\ck$ \cite{pol}.

In the appendix we will discuss the 1-loop amplitudes to some extent.
We begin with the Klein bottle whose structure determines the orientifold
planes and the type of D-branes to be added. Furthermore,
the difference between $\Omega$ and $\op$ manifests only in $\ck$.
The cylinder and M\"obius strip amplitudes will also be explained.

{} From the one-loop amplitudes we can deduce the spectrum of tachyons and
massless states. General properties of closed and open string states
will be described in sections \ref{sscc} and \ref{ssca}.

\subsection{Non-supersymmetric $\Z_N$}
\label{sszn}

To describe the action of the $\Z_N$ generator $\theta$ on bosonic and fermionic movers
$X^M$, $\psi^M$ in the light cone ($M=2,\cdots, 9$), it is
convenient to use complex basis $Y^a = X^{2a+2} + i X^{2a+3}$ and
$\Psi^a = \psi^{2a+2} + i \psi^{2a+3}$, $a=0, \cdots, 3,$ in which $\theta$ is diagonal,
i.e. $\theta Y^a=e^{2i\pi v_a}Y^a$ and $\theta \Psi^a=e^{2i\pi v_a}\Psi^a$,
where $N v_a \in \Z$ because $\theta^N=\uno$.
Space-time is $({\rm Minkowski})^D$ with  $D=10-2d$, and the internal space
is the orbifold $\T^{2d}/\Z_N$, with coordinates $Y^i$, $i=4-d, \cdots, 3$. In
the cases of interest, $d\leq 3$. Notice that Lorentz invariance requires
$v_a \in \Z$ for $a=0,\cdots, 3-d$.
The $\T^{2d}$ lattice is denoted $\Lambda$. Recall that windings $W$ take values in
$\Lambda$, whereas quantized momenta $P$ belong to the dual $\Lambda^*$. Under
$\Omega$, $W \to -W$ and $P \to P$.

Upon compactification,
the little group $\so(8)$ breaks to $\so(D\!-\!2) \times \so(2d)$ so that string states
are classified in terms of $\so(D\!-\!2)$ representations. For example,
in the right Neveu-Schwarz (NS) sector of closed strings the massless states
$\psi_{-\oh}^M | 0 \rangle$ form an  $\so(8)$ vector with
weights ${\bf 8}_v=(\underline{\pm{}1,0,0,0})$, where underline stands for permutations.
Under $\so(D\!-\!2)$,  ${\bf 8}_v$ branches into a vector and $d$ complex scalars
$\Psi_{-\oh}^i | 0 \rangle$. In the Ramond (R) sector, the
vacuum is massless and is given by an $\so(8)$ spinor with weights
${\bf 8}_s=\pm(\underline{-\oh,\oh,\oh,\oh})$.
In general, each state has a weight $r$ of $\so(8)$. The
GSO projection is $\sum_a r_a={\rm odd}$. Besides, the action of $\theta$
is simply $\theta \ket{r} = e^{2i\pi r \cdot v} \ket{r}$, where
$v=(v_0,v_1,v_2,v_3)$ is the twist vector.

The $v_a$'s are constrained by the condition that $\theta^m$ must act
crystallographically on the lattice $\Lambda$ of the internal torus.
In addition, modular invariance imposes
\beq
N S_v={\rm even} \quad ; \quad S_v \equiv \sum_a v_a \  ,
\label{modinv}
\eeq
which in turn ensures that $\theta$ is of order $N$
acting on the world-sheet fermionic degrees of freedom .
Supersymmetry requires the existence of invariant states
$| r \rangle$, with $r$ an spinorial weight in ${\bf 8}_s$.
For $v_0=0$ this gives the well known condition
\beq
\pm v_1 \pm v_2 \pm v_3 = 0 \, {\rm mod} \, 2  .
\label{susycon}
\eeq
Relaxing this condition breaks supersymmetry.
The allowed non-supersymmetric  twist vectors
for $N\leq 6$ were found in \cite{fh}. They are:
\beq
\renewcommand{\arraystretch}{1.25}
%\begin{center}
%\begin{tabular}{|c|c||c|c|}
\begin{array}{|c|l||c|l|}
\hline
%\Z_N & (v_0,v_1,v_2,v_3) & \Z_N & (v_0,v_1,v_2,v_3)  \\
%\hline
\Z_2^* & (0,0,0,1) &
\Z_6^* & (0,0,0,\frac13)  \\
\cline{1-2}
\Z_3^* & (0,0,0,\frac23)  &
{} & (0,0, \frac13, \frac23)  \\
\cline{1-2}
\Z_4 & (0,0, 0, \frac12)  &
{} & (0,\frac13, \frac13, \frac13) \\
\cline{3-4}
{} & (0,0, \frac14, \frac34)  &
\Z_6 & (0,0, \frac16, \frac12)   \\
{} & (0,\frac12, \frac12, \frac12)&
{} & (0,0, \frac16, \frac56)  \\
\cline{1-2}
\Z_5^* & (0,0, \frac15, \frac35)  &
{} & (0,\frac23, \frac16, \frac16)   \\
{} & {}  &
{} & (0,\frac13, \frac12, \frac12) \\
\hline
\end{array}
%\end{tabular}
%\end{center}
%\caption{Twist vectors for non-supersymmetric $\Z_N$ actions, $N \leq 6$.}
\label{tab1}
%\end{table}
\eeq
For larger $N$ there are many more allowed twists. For the torus
lattice for each action we will take products of two-dimensional
sub-lattices whenever allowed. Concretely, for order two and order
four rotations we take the SO(4) root lattice whereas for order
three and order six rotations we take the SU(3) root lattice. The
$\Z_5$ action is realized on the SU(5) root lattice.

In the examples denoted $\Z_N^*$, $\theta^m Y^i \not= -Y^i$, so that
$\Omega \theta^m$ leaves no sub-lattice invariant and the resulting
orientifolds will contain only O9-planes. The $\Z_2^*$ gives actually
an orientifold of type 0B in \deq10 since $v_3$ is integer and
in fact $\theta=(-1)^{F_S}$. Notice also that in the $\Z_6^*$,
$3v_a$ is an integer and $\theta^3=(-1)^{F_S}$.

For the remaining actions in (\ref{tab1}) there will also be lower dimensional O-planes.
Concretely, in all $\Z_6$ and in the $\Z_4$ in \deq6, there are ${\rm O}5_1$-planes,
with $Y^2$ and $Y^3$ transverse, because the element $\theta^\nm$ reflects
these two coordinates and leaves $Y^1$ invariant. The $\Z_4$ in \deq8 has ${\rm O}7_3$-planes,
with $Y^3$ transverse, since $\theta Y^3 = -Y^3$. Finally, the $\Z_4$ in \deq4
will include O3-planes because $\theta$ reflects all internal coordinates.

\subsection{Closed string states}
\label{sscc}

Modular invariance requires the existence of sectors twisted by
$\theta^n$, $n=0, \cdots, N-1$. In each sector the states are
tensor products $\ket{R} \times \ket{L}$ of right and left modes.
In turn, $\ket{R}=\ket{N_R, r_{nR}}$, where $N_R$ is an oscillator number and
$r_{nR}=r_R + nv$ with $r_R$ an $\so(8)$ weight (similar for $\ket{L}$). To determine
the states in the orientifold spectrum, we start with invariant combinations under $\Z_N$.
For example, in the untwisted sector it suffices to have
$(r_R-r_L)\cdot v={\rm integer}$ for states without oscillators acting on them.
In the twisted sectors we must take into account the structure of fixed points
that depends on the torus lattice, see the appendix of \cite{fh} for more details.

We then implement the orientifold projection, i.e. invariance under $\Omega_P$.
In the untwisted sector both $\Omega$ and $\op$ just exchange left and right modes.
Then, the invariant combinations must be symmetric in the NSNS and  [NSR + RNS] sectors,
but antisymmetric in the RR sector since fermionic modes are exchanged.
In the $\theta^\nm$ sector, when $N$ is even, the invariant combinations are the
same for $\Omega_P=\Omega$. On the other hand, as explained in section \ref{sskb}, 
for $\Omega_P=\op$, when $\frac{N}2 v$ is such that $\theta^\nm$ is equivalent
to $(-1)^{F_S}$,
there is an extra minus sign in the Klein-bottle amplitude so that in the $\theta^\nm$
sector we need take the opposite type of combinations as in the untwisted sector.
For the remaining twisted sectors we have to take into account that
$\theta^n \rightarrow \theta^{N-n}$ under $\Omega_P$.

\subsection{Open string states}
\label{ssca}

Including labels $ab$ for the ends on $\D p$ and $\D q$-branes,
states are of the form
\beq
\ket{\phi,ab} (\lambda^{\phi}_{pq})_{ab} \  ,
\label{oss}
\eeq
where $\lambda^{\phi}_{pq}$ is the Chan-Paton matrix and $\phi$ represents
the world-sheet modes. The action of $\theta^m$ and $\Omega_P \theta^m$ on the
Chan-Paton  matrices is realized by the unitary matrices $\gamma_{m,p}$ and
$\gamma_{\Omega m,p}$ such that
\beq
\theta^m : \lambda^{\phi}_{pq} \rightarrow
\gamma_{m,p}\lambda^{\phi}_{pq}\gamma_{m,q}^{-1}
\quad ; \quad
\Omega_P \theta^m:
\lambda^{\phi}_{pq} \rightarrow
\gamma_{\Omega m,q}{\lambda^{\phi}_{pq}}^T \gamma_{\Omega m,p}^{-1}
\  ,
\label{gaction}
\eeq
since $\Omega_P$ exchanges the ends.

The $\gamma$ matrices form a representation of $\cg$ up to a phase \cite{gp}.
Consistent with group multiplication we can define
\beq
\gamma_{m,p}= \gamma_{1,p}^m \quad ; \quad
\gamma_{\Omega m,p}= \gamma_{m,p} \gamma_{\Omega,p}  \  .
\label{gdefs}
\eeq
It can be shown that $\gamma_{0,p}=\uno$ \cite{gp}. Besides, $\theta^N=\uno$,
gives
\beq
\gamma_{N,p}=(\gamma_{1,p})^N=\delta_p \uno  \ ,
\label{deltadef}
\eeq
where $\delta_p=\pm 1$, keeping the sign freedom. 

When only D9 and D5-branes are present, it happens that
$(\Omega_P\theta^m)^2=\theta^{2m}$ acting on $pp$ strings.
It then follows that 
\beq
\gamma_{\Omega m,p}= \epsilon_{m,p} \, \gamma_{2m,p} \gamma_{\Omega m,p}^T  \  ,
\label{epsilondef}
\eeq
where $\epsilon_{m,p}=\pm 1$.
Now, we can always absorb a phase in $\gamma_{1,p}$ to attain
\beq
\epsilon_{m,p}=\epsilon_{0,p} \equiv \epsilon_p  \ .
\label{epsilonsim}
\eeq
It is then useful to recast (\ref{epsilondef}) as
\beq
\gamma_{m,p}^* = \epsilon_p \gamma_{\Omega,p}^*
\gamma_{m,p} \gamma_{\Omega,p}  \  .
\label{gmcon}
\eeq
The $\gamma$ matrices are further constrained by tadpole
cancellation conditions. For antibranes, $\delta_{\bar
p}=\delta_p$ and $\epsilon_{\bar p}=\epsilon_p$.
In section \ref{seco9o73} we will explain how these results are modified
when D3 or D7-branes are included.

To understand the meaning of $\epsilon_p$, first use (\ref{epsilondef})
to derive the relations
\beq
\gamma_{\Omega ,p}^T= \epsilon_p \, \gamma_{\Omega ,p}
\quad ; \quad
\Tr \inpar{\gamma_{\Omega ,p}^{-1} \gamma_{\Omega ,p}^T}= \epsilon_p N_p
\ ,
\label{epzero}
\eeq
where $N_p$ is the total number of $\D p$-branes.
For instance, when $\epsilon_p=1$ we can take $ \gamma_{\Omega ,p}=\uno$,
leading to orthogonal projection $\lambda^T=-\lambda$ on the Chan-Paton matrix
of $pp$ gauge vectors (see below). For $\epsilon_p=-1$ there is instead a
symplectic projection. The second relation in (\ref{epzero}) means that
$\epsilon_p$ correlates with the sign of the tension and the RR charge
of ${\rm O}p$-planes. In particular, $\epsilon_p=1$ corresponds to
${\rm O}p$-planes of negative tension and RR charge. For $\epsilon_p=-1$ both
signs are reversed. In the notation of \cite{as} these are ${\rm O}p+$ and ${\rm O}p-$
respectively.

To determine the world sheet modes $\phi$, and then find the states invariant under $\cg$, \
it is necessary to specify the type of branes at the ends.
For states coming from open strings ending at the same type of branes we
need to look at one-loop cylinder and M\"obius strip amplitudes  $\cc_{pp}$ and
$\cam_p$, as well as $\cc_{\bar p \bar p}$ and $\cam_{\bar p}$ if there are
antibranes. For open strings ending at different branes it suffices to
study $\cc_{pq}$, $\cc_{\bar p \bar q}$ and  $\cc_{p \bar q}$. Since $\Omega_P$
exchanges the ends, these open strings cannot enter in M\"obius strip traces.
We give explicit results for  configurations with $\D 9 + \ov{\D 9}$ that are
generically present, and also with $\D 5 + \ov{\D 5}$ that appear in several
$\Z_{even}$ examples. There are cases with $\D 7 + \ov{\D 7}$ or
$\D 3 + \ov{\D 3}$ that will be explained when discussing the concrete examples.

\subsubsection{$99$, $\bar{9}\bar{9}$ and $9\bar{9} + \bar{9}9$ states}
\label{sss99}

For 99 strings, $\phi$ is given by the right modes of the closed string.
To each state we assign an $\so(8)$ weight $r$ verifying the
GSO projection $\sum_a r_a ={\rm odd}$ that eliminates the
tachyon and leaves massless states
$r={\bf 8}_v$ in the NS sector and  $r={\bf 8}_s$ in the R sector.
The Chan-Paton matrix $\lambda^{r}_{99}$ of the massless invariant states
must satisfy
\beq
\lambda^{r}_{99} = e^{2i \pi  r \cdot v}
\gamma_{1,9}\lambda^{r}_{99}\gamma_{1,9}^{-1} \quad , \quad
\lambda^{r}_{99} = -\gamma_{\Omega,9}{\lambda^{r}_{99}}^T
\gamma_{\Omega,9}^{-1} .
\label{enn}
\eeq
For $\bar{9}\bar{9}$ strings, the cylinder partition function is identical to that of
99 strings, thus the massless weights are the same. The invariance conditions
are analogous to (\ref{enn}) except for an extra minus sign in the
$\Omega_P$ projection for the R states. This is due to the sign change in
the R sector of the M\"obius strip amplitude $\cam_{\bar 9}$, which in turn
corresponds to the opposite RR charge of $\ov{\D 9}$-branes.

The  cylinder amplitude $\cc_{9\bar 9}$ is obtained from
$\cc_{99}$ by reversing the relative sign between the two
NS (and the two R) contributions. We can still assign $\so(8)$ weights to the
various states but the GSO projection changes to $\sum_a r_a ={\rm even}$.
The NS sector then contains a tachyon with $r=0$ and no massless states.
In the R sector there is a massless spinor of opposite chirality, with weights
$r={\bf 8}_c=\pm(\oh,\oh,\oh,\oh), (\underline{-\oh,-\oh,\oh,\oh})$.
Invariance under $\theta$ implies:
\beq
\lambda^{r}_{9\bar{9}}=e^{2i \pi r \cdot v}
\gamma_{1,9}\lambda^{r}_{9\bar{9}}\gamma_{1,\bar{9}}^{-1} .
\label{ennb}
\eeq
Under $\Omega_P$, $9\bar{9} \to \bar{9}9$, so that we must just retain
half of the states in the spectrum without further restrictions.

To find the states arising under compactification we just decompose the
$\so(8)$ weights under  $\so(D\!-\!2) \times \so(2d)$. From ${\bf 8}_v$
we clearly obtain a vector, with $r\cdot v=0$, plus scalars with
$r\cdot v=\pm v_i$. In the 99 and $\bar 9 \bar 9$ sectors the
fermions come from ${\bf 8}_s$ which in the different dimensions decomposes as
\beqa
D=8 & : & \ \ \  \r8_s = (\r4, -\ts{\frac12}) + (\bar{\r4},\ts{\frac12}) \ ,
\nonumber \\[0.2cm]
D=6 & : & \ \ \  \r8_s = (\r2_{\rm L}, \r2_{\rm R}) + (\r2_{\rm R},\r2_{\rm L}) \ ,
\label{branch8} \\[0.2cm]
D=4 & : & \ \ \  \r8_s = (\ts{\frac12}, \bar{\r4}) + (-\ts{\frac12},\r4)  \  .
\nonumber
\eeqa
We use conventions such that for $\so(6)$, $\r4=(\oh,\oh,\oh), (\underline{-\oh,-\oh,\oh})$.
Notice that the $\so(2)$ charge in \deq8 is the component $r_3$ of $r$,
whereas in \deq4 it is $r_0$ which corresponds to helicity.
For the $\so(4)$ representations in \deq6 our convention is
$\r2_{\rm L}=\pm(\oh,-\oh)$ and $\r2_{\rm R}=\pm(\oh,\oh)$.
In $D\!=\!8,4$ we will distinguish fermions from antifermions by the
condition $\prod_{i=(D-2)/2}^3 r_i < 0$.

When the rotation breaks supersymmetry, none of the 99 or $\bar 9 \bar 9$
fermions has weights with $r\cdot v=0$. Thus, there are no gauginos to
pair up with gauge vectors. Instead, there are charged fermions whose $r\cdot v$
follows from the branching rules in (\ref{branch8}). For example, in
\deq8 the fermions belong to the $\r4$ of $\so(6)$ and have $r\cdot v=-v_3/2$
(antiparticles have the opposite). In \deq4 the fermionic particles have
helicity $1/2$ and come in four types according to $r\cdot v=-S_v/2$ or
$r\cdot v=-v_i+S_v/2$. In \deq6 there are fermions in $\r2_{\rm L}$ with
$r\cdot v=S_v/2$ and in $\r2_{\rm R}$ with $r\cdot v=-v_3+S_v/2$, plus
the complex conjugates. In the $9 \bar 9$ sector the \deq10 massless spinor
is $\r8_c$. Hence, the fermions arising from this sector will have opposite
chiralities.

\subsubsection{$55$, $\bar{5}\bar{5}$ and $5\bar{5} + \bar{5}5$ states}
\label{sss55}

For concreteness we focus on $\D 5_1$-branes. Then the coordinate
$Y^1$ has NN whereas $Y^{2,3}$ have DD boundary conditions. Since
there are no mixed boundary conditions, the massless 55 states can
be labelled by $\so(8)$ weights, $\r8_v$ in NS and $\r8_s$ in R,
that must be appropriately decomposed upon compactification. It is
necessary to distinguish the states associated to directions with
DD or NN boundary conditions. For $\D 5_1$-branes the
corresponding $r$'s are
\beqa
r_v^{\rm DD} = (0,0,\underline{\mp 1, 0}) \quad & ; & \quad
r_s^{\rm DD} = \pm (\ts{\oh},\ts{\oh},\underline{-\ts{\oh},
\ts{\oh}}) \nonumber \\[0.2cm]
r_v^{\rm NN} = (\underline{\mp 1, 0},0,0) \quad & ; & \quad
r_s^{\rm NN} = \pm (\underline{-\ts{\oh},\ts{\oh}},
\ts{\oh},\ts{\oh}) \ .
\label{r55}
\eeqa
This is valid when $\frac{N}2 S_v={\rm even}$, otherwise $r_s^{\rm
DD}$ and $r_s^{\rm NN}$ are exchanged. The point is that in
supersymmetric $\Z_N$ some or all of the $r_s^{\rm NN}$ correspond
to gauge spinors that must satisfy $r_s^{\rm NN} \cdot v \in \Z$
in order to pair with gauge vectors having $r_v^{\rm NN} \cdot
v=0$.

The full orientifold projection requires that the Chan-Paton
factors $\lambda_{55}^r$ satisfy
\beq
\lambda^{r}_{55} = e^{2i \pi  r \cdot v}
\gamma_{1,5,J}\lambda^{r}_{55}\gamma_{1,5,J}^{-1} \quad , \quad
\lambda^{r}_{55} = \pm \gamma_{\Omega,5}{\lambda^{r}_{55}}^T
\gamma_{\Omega,5}^{-1} \ .
\label{eff}
\eeq
The index $J$ in $\gamma_{1,5,J}$ refers to the fixed point of
$\theta$ where the $\D 5_1$-branes are located. When the branes
sit at a point in the bulk there is no orbifold projection since
$\theta$ only exchanges images. In the $\Omega$ projection, the
plus sign applies to states associated with DD boundary
conditions. Notice that for fermions, DD or NN is correlated with
the chirality of the spinor in the brane world-volume.
If the branes sit at a fixed point of $\theta^m$, but not of $\theta$,
in (\ref{eff}) we need replace $v \to mv$ and
$\gamma_{1,5,J} \to \gamma_{m,5,J}$. 

For $\bar 5 \bar 5$ states the massless weights are the same as in
(\ref{r55}). The projection is similar to (\ref{eff}) except for
an extra minus sign, due to the opposite RR charge, in the
$\Omega$ projection of R states.

In the $5 \bar 5$ sector the GSO projection allows instead a
tachyon with $r=0$ and a massless spinor $\r8_c$ of opposite
chirality. We only need to impose the $\theta$ projection
\beq
\lambda^{r}_{5\bar 5} = e^{2i \pi  r \cdot v}
\gamma_{1,5,J}\lambda^{r}_{5\bar 5}\gamma_{1,\bar 5,J}^{-1} \ .
\label{effb}
\eeq
The $\Omega$ projection just exchanges states in $5\bar 5$ with
those in $\bar 5 5$.

\subsubsection{$95 + 59$, $\bar 9\bar{5} + \bar{5}\bar 9$,
$9\bar{5} + \bar{5}9$ and $\bar 9 5 + 5 \bar 9$ states}
\label{sss95}

It is enough to discuss 95 states in some detail. The new feature
is the appearance of mixed boundary conditions. For $\D
5_1$-branes, $Y^{0,1}$ are NN but $Y^{2,3}$ are ND. As explained
in appendix A, the partition function changes in such a
way that massless states in the NS and R sector have weights
\beq
r_{\rm NS}=(0,0, w_2, w_3) \quad ; \quad r_{\rm R}=(w_0, w_1, 0,
0)  \ .
\label{r95}
\eeq
The $w_a$ take values $\pm \oh$ and are constrained by the GSO
projection that depends on whether $\frac{N}2 S_v$ is even or odd.
Concretely, for $\frac{N}2 S_v$ even, $\sum_a w_a={\rm odd}$, but
for $\frac{N}2 S_v$ odd, $\sum_a w_a={\rm even}$. The $\theta$
projection is simply
\beq
\lambda^{r}_{95} = e^{2i \pi  r \cdot v}
\gamma_{1,9}\lambda^{r}_{95}\gamma_{1,5,J}^{-1} \ .
\label{enf}
\eeq
The $\Omega$ projection only effects $95 \leftrightarrow 59$.

For $\bar 9 \bar 5$ states the massless weights are the same as
for 95. For $\bar 9 5$ and $9 \bar 5$ the massless weights have
the same form as in (\ref{r95}) but the GSO projection is the
opposite compared to 95. In all cases the $\theta$ projection is
analogous to (\ref{enf}).

\section{Models with O9-planes}
\label{seco9}

This is the case of the $\Z_N^*$ actions whose
elements do not invert any coordinate.
To begin we use the results of the appendix to
deduce the tadpole cancellation conditions.
We then determine the $\gamma$ matrices that form a representation of $\cg$ and
cancel RR tadpoles.  As explained before, the group structure
allows a freedom in $\epsilon_{0,p}=\epsilon_p$ that enters in (\ref{gmcon}).
We will consider both values for $\epsilon_p$. In several examples
we will present the resulting spectrum of massless and tachyonic states
for both $\Omega$ and $\Omega^\prime$ projections.

An useful check on the spectrum is the cancellation of the
irreducible gauge anomaly proportional to $\tr F^{\frac{D+2}2}$
for each group factor. In $D\!=\!10,6$,  the irreducible
gravitational anomaly must vanish as well. In \deq6, cancellation
of the $\tr R^4$ anomaly requires
\beq
N_L - N_R = 28(n_- - n_+)  \  ,
\label{ar4}
\eeq
where $n_\pm$ is the number of tensors $\r3^\pm$ whereas $N_L$ and
$N_R$ are the numbers of $2(\r2_{\rm L})$ and $2(\r2_{\rm R})$
respectively.

\subsection{Tadpole cancellation with O9-planes}
\label{sstco9}

For the $\Z_N^*$ actions there are only untwisted tadpoles
proportional to  a 10-dimensional volume $V_{10}$, and twisted
tadpoles proportional to $V_D$. The reason is that
the invariant momentum sub-lattice is either trivial
and its volume is $V_P=1$ by definition, or it is the full $\Lambda^*$
and $V_DV_P$ is $V_{10}$. We are employing the
notation introduced in the appendix.

Let us first consider the divergences from the Klein bottle amplitude. 
When $\Omega_P=\Omega$, the RR tadpole (\ref{tadkbrrfull}) simplifies to 
\beq T_\ck^{\rm RR}(n,m) = e^{-i\pi n S_v} 2^{D+I_P} V_D V_P \!
\prod_{2mv_j\notin\Z}  \! \! \! 2 \inmod{\sin 2\pi mv_j}  \  . 
\label{tadkbrr} 
\eeq 
For the NSNS tadpoles there is an analogous formula but with the phase 
$e^{-i\pi n S_v}$ absent. When $N$ is even, the RR tadpoles vanish since 
\beq 
\sum_{m=0}^{N-1} \incor{T_\ck^{\rm RR}(0,m) + T_\ck^{\rm RR}(\ts{\frac{N}2},m)}=0 \ . 
\label{krr1}
\eeq 
However, there are left-over NSNS tadpoles, either proportional to $V_{10}$ or to $V_D$, that can be
cancelled by introducing D9-branes. The RR tadpoles created by the D9-branes can in turn be cancelled by adding
$\ov{\rm D9}$-branes. For $N$ odd adding D9-branes is mandatory since there are uncancelled RR tadpoles.

When $\Omega_P=\Omega^\prime$, there is an extra minus sign in $T_\ck^{\rm RR}(\frac{N}2,m)$.
Thus, the divergences from the untwisted and the $\frac{N}2$-twisted
sector add to produce a non-zero RR tadpole. On the other hand, the NSNS tadpoles
from both sectors cancel each other.

{}From the $\cc_{99}$ cylinder the RR tadpole (\ref{tadrrpp}) reduces to:
\beq
T_{99}^{\rm RR}(m) =
\inpar{{\rm Tr}  \gamma_{m,9}}^2  V_D  V_P \!
\prod_{mv_j\notin\mathbb{Z}} \! 2 \inmod{\sin \pi mv_j}.
\label{tadrr99}
\eeq
In the presence of ${\ov{\D 9}}$-branes, we need to substitute
$\Tr \, \gamma_{m,9}$ by  $\Tr \, \gamma_{m,9} - \Tr \, \gamma_{m,\bar 9}$.

In $\cam_9$ we find:
\beq
T_{9}^{\rm RR}(m)  =  -
2^{\oh(D+I_P+2)} \inpar{\epsilon_9\, \Tr \gamma_{2m,9}}
V_D V_P  \prod_{a=0}^3 c(mv_a) \,
\prod_{mv_j\notin\Z} 2 \inmod{\sin \pi{}mv_j}
\label{tadrrm9}
\eeq
where we have used (\ref{epsilondef}).
For  $\cam_{\bar 9}$ there is an extra minus sign due to the
opposite RR charge.

Clearly, the tadpoles (\ref{tadkbrr}), (\ref{tadrr99}) and  (\ref{tadrrm9})
have a definite volume dependence. In fact, they are proportional either to
$V_{10}$ or to $V_D$.
Furthermore, to derive the cancellation conditions the tadpoles from the
various amplitudes are combined according to the twisted sector in
the transverse (tree-level) channel. For $N$ odd and $\forall m$ the
RR tadpoles cancel provided that
\beq
\Tr  \gamma_{2m,9} - \Tr  \gamma_{2m,\bar 9} = 32 \, \epsilon
\prod_{j=1}^{d} \cos m\pi v_j \ ,
\label{tconimpar}
\eeq
where $\epsilon\equiv \epsilon_9 = \epsilon_{\bar 9}$.
Additionally, we can choose $\gamma_{N,p}=\uno$
($p=9, \bar 9$) without loss of generality.
In case of orthogonal projection, $\epsilon=1$, we can avoid the
$\ov{\D 9}$-branes. For the symplectic projection, $\epsilon=-1$,
the O9-planes have positive RR charge and $\ov{\D 9}$-branes are
necessary. Concerning NSNS tadpoles, for the $\Z_3$ and $\Z_5$ in
(\ref{tab1}) we have found that they vanish when the RR tadpoles do.

Let us now consider $N$ even. For $\Omega_P=\Omega$,
cancellation of RR tadpoles in $\ck$ shows the existence
of two types of orientifold planes with opposite charges. This interpretation
is possible because in $\Z_N^*$ orientifolds there are two sets of RR
potentials, one set from the untwisted sector and the other from the
$\theta^\nm$ sector. By consistency, the RR M\"obius tadpoles
must also cancel. This requires $\gamma_{N, p}= \uno$, for $p=9, \bar 9$.
Additionally, to cancel the cylinder tadpoles,
\beq
\Tr \gamma_{m,9} - \Tr \gamma_{m,\bar 9} = 0 \quad , \quad \forall m  \ .
\label{tconpar}
\eeq
This is the RR tadpole cancellation condition in a type 0B orientifold on
$\T^{2d}/\tilde \Z_{\frac{N}2}$, with $\frac{N}2$ odd and
$\tilde v=(v_0, v_1, v_2, v_3 -1)$. As for NSNS tadpoles, adding the
pieces from the different amplitudes we obtain the cancellation
condition
\beq
\Tr  \gamma_{2m,9} + \Tr \gamma_{2m,\bar 9} = 64 \, \epsilon
\prod_{j=1}^{d} \cos m\pi \tilde v_j  \  .
\label{tnsns}
\eeq
Notice that for $\epsilon=-1$ the NSNS tadpoles cannot be cancelled
at all. This is as expected\, since in this case ${\rm O}9$-planes have positive tension
as $\D 9$-branes and  $\ov{\D 9}$-branes do. However, for $\epsilon=1$,
both RR and NSNS tadpoles can vanish as we will exemplify later.

We finally come to $N$ even and $\Omega_P=\Omega^\prime$. Now
there is a RR tadpole from the Klein bottle and necessarily from the M\"obius
amplitude. Then, it must be $\gamma_{N, p}= -\uno$, for $p=9, \bar 9$.
Collecting all pieces we obtain the RR tadpole cancellation conditions
\beqa
\Tr  \gamma_{2m+1,9} - \Tr \gamma_{2m+1,\bar 9} & = & 0
\nonumber \\[0.2cm]
\Tr \gamma_{2m,9} - \Tr \gamma_{2m,\bar 9} & = & 64 \,  \epsilon
\prod_{j=1}^{d} \cos m\pi  v_j \  ,
\label{tparop}
\eeqa
both for $m=0, \cdots, \frac{N}2-1$. There are always NSNS tadpoles
left uncancelled.

\subsection{Solutions with $\Omega$ projection}
\label{O9Omega}

As we have seen in \ref{sstco9}, for $N$ even necessarily $\gamma^N_{1,p}=\uno$
($p=9,\bar 9$) and for $N$ odd we can make the same choice without altering the results.
Then, in general ($\mu=e^{2i\pi/N}$)
\beq
\gamma_{1,9} = {\rm diag}(\uno_{n_0}, \mu \uno_{n_1},\mu^2 \uno_{n_2}, \cdots, \mu^{N-1} \uno_{n_{N-1}})
\quad ; \quad
n_{N-j}=n_j  \ ,
\label{gonestaro}
\eeq
where $\uno_n$ is the $n\times n$ identity matrix. For $\gamma_{1,\bar 9}$, replace $n_j$
by $\bar n_j$. Imposing RR tadpole cancellation relates the $n_j$ to the $\bar n_j$. For
instance, for $N=2L+1$, (\ref{tconimpar}) becomes
\beq
(n_0 - \bar n_0) + 2 \sum_{k=1}^L (n_k - \bar n_k) \cos \ds{\frac{4\pi mk}N}
= 32 \, \epsilon
\prod_{j=1}^{d} \cos(m\pi v_j)
\quad ; \quad m=0, \cdots, L   \  .
\label{nconodd}
\eeq
For $N=2L$, (\ref{tconpar}) simply gives $n_j = \bar n_j$. To cancel NSNS tadpoles, according
to (\ref{tnsns}), there are further conditions
\beq
n_0 + n_L + 2 \sum_{k=1}^{L-1} n_k  \cos \ds{\frac{4\pi mk}N}
= 32 \, \epsilon
\prod_{j=1}^{d} \cos(m\pi \tilde v_j)
\quad ; \quad m=0, \cdots, L-1   \  .
\label{nconeven}
\eeq
Clearly, this needs $\epsilon=1$.

Concerning the realization of $\Omega$, in agreement with (\ref{gmcon}), for $\epsilon_9 \equiv \epsilon=1$
we can take
\beq
\gamma^+_{\Omega,9}=\inpar{\!\!\begin{array}{cccccc}
\uno_{n_0} & 0 & 0 & \cdots & 0 & 0\\
0 & 0 & 0 & \cdots & 0 & \uno_{n_1}\\
0 & 0 & 0 & \cdots & \uno_{n_2} & 0 \\
\vdots & & & & & \vdots \\
0 & 0 & \uno_{n_2} & \cdots & 0 & 0 \\
0 & \uno_{n_1} & 0 & \cdots & 0 & 0
\end{array}\!\!}  \  .
\label{gomegaort}
\eeq
For $\epsilon=-1$ we need to distinguish between $N$ odd or even. For $N=2L+1$,
\beq
\gamma^-_{\Omega,9}=\inpar{\!\!\begin{array}{ccccccc}
iJ_{n_0} & 0 & \cdots & 0 & 0 &  \cdots  & 0\\
0 & 0 & \cdots & 0 & 0 & \cdots & i\uno_{n_1}\\
\vdots & & & & & & \vdots \\
0 & 0  & \cdots & 0 & i\uno_{n_L} & \cdots &  0 \\
0 & 0  & \cdots & -i\uno_{n_L} & 0 & \cdots &  0 \\
\vdots & & & & & & \vdots \\
0 & -i\uno_{n_1} & \cdots & 0 & 0 & \cdots & 0
\end{array}\!\!}
\quad ; \quad
J_{n_0}=\inpar{ \! \! \begin{array}{cc}
0&\uno_{n_0/2}\\
-\uno_{n_0/2}&0\\
\end{array}  \! \!} \  .
\label{gomegasympodd}
\eeq
Clearly $n_0$ must be even. Instead, for $N=2L$,
\beq
\gamma^-_{\Omega,9}=\inpar{\!\!\begin{array}{cccccccc}
iJ_{n_0} & 0 & \cdots & 0 & 0 & 0&  \cdots  & 0\\
0 & 0 & \cdots & 0 & 0 & 0& \cdots & i\uno_{n_1}\\
\vdots & & & & & & & \vdots \\
0 & 0  & \cdots & 0 & 0 &  i\uno_{n_{L-1}} & \cdots &  0 \\
0 & 0  & \cdots & 0 & iJ_{n_L} & 0& \cdots &  0 \\
0 & 0  & \cdots & -i\uno_{n_{L-1}} & 0 & 0 & \cdots &  0 \\
\vdots & & & & & & & \vdots \\
0 & -i\uno_{n_1} & \cdots & 0 &  0 & 0 & \cdots & 0
\end{array}\!\!} \  ,
\label{gomegasympeven}
\eeq
with both $n_0$ and $n_L$ even.

Consider now the the open string spectrum.
The task is to determine the Chan-Paton factors after substituting the above
$\gamma$ matrices in, say (\ref{enn}) for 99 states. The simplest case is
that of gauge vectors that have $r\cdot v=0$. From the form of the resulting
$\lambda$'s we can read the gauge groups. For $\epsilon=1$ we find
\beqa
& & N=2L+1 \ \ \  : \ \ \  G_9= \so(n_0) \times \u(n_1) \times \cdots \times \u(n_L)  \  ,
\nonumber \\[0.2cm]
& & N=2L \ \ \ \ \ \ \ \ \, : \ \ \
G_9= \so(n_0) \times \u(n_1) \times \cdots \times \u(n_{L-1}) \times \so(n_L)  \  .
\label{g99om}
\eeqa
The $\bar 9 \bar 9$ group factors in $G_{\bar 9}$ are the same
with $n_k \to \bar n_k$. For $\epsilon=-1$, we just need replace
$\so(n) \to \us(n)$ and take $n$ even.

The massless charged scalars and fermions transform under $\theta$ as 
explained in section \ref{sss99}.
Given the possible values of $r\cdot v$ we find the Chan-Paton matrices
according to (\ref{enn}) and then read off the corresponding representations.
Specific examples will be presented in section \ref{ssecexamples}.
In the $9\bar 9$ sector besides
massless charged fermions there are charged tachyons that have $r=0$. Under
$G_9 \times G_{\bar 9}$ they turn to transform as
\beq
\inpar{\repf_0,\repf_{\bar{0}}} + \sum_{k=1}^{\left[\frac{N-1}2 \right]} \,
\incor{\inpar{\repf_k,\repfc_{\bar k}}+ {\rm c.c.} }
+ \inpar{\repf_{\frac{N}2},\repf_{\bar{\frac{N}2}}}  \  ,
\label{tachy}
\eeq
where the last term appears only if $N$ is even. The subscripts attached
to the Young tableaux refer to the $k$-th group factor in $G_9$ or in $G_{\bar 9}$.
It is understood that the states are singlets under absent factors.

\subsection{Solutions with $\Omega^\prime$ projection}
\label{O9Omegap}

As explained  in \ref{sstco9} we only need consider $N$ even ($N=2L$)
and $\gamma^N_{1,p}=-\uno$. Then, in general ($\nu=e^{i\pi/N}$)
\beq
\gamma_{1,9} = {\rm diag}(\nu\uno_{n_1}, \nu^3 \uno_{n_2},\cdots, \nu^{2N-1} \uno_{n_N})
\quad ; \quad
n_{N-j+1}=n_j  \ .
\label{gonestarop}
\eeq
The $\gamma_{1,\bar 9}$ is analogous with $n_j \leftrightarrow \bar n_j$.
There are relations between the dimensions $n_j$ and $\bar n_j$ following from
the cancellation conditions (\ref{tparop}). For instance, for the $\Z_2^*$,
$(n_1 - \bar n_1)=32 \epsilon$, whereas for the
$\Z^*_6$ in (\ref{tab1}),
\beq
(n_1 - \bar n_1) = (n_3 - \bar n_3) = \frac{32}3 \epsilon (1 + \ts{\prod_j} \cos \pi v_j)
\quad ; \quad
(n_2 - \bar n_2) = \frac{32}3 \epsilon (1 - 2\ts{\prod_j} \cos \pi v_j)   \  .
\label{z6sdims}
\eeq
Notice that we can take $\epsilon=-1$ only if $\ov{\D 9}$-branes are present.

The realization of $\Omega^\prime$ is given by
\beq
\gamma^+_{\Omega,9}=\inpar{\!\!\begin{array}{ccccc}
0 & 0 & \cdots & 0 & \uno_{n_1}\\
0 & 0 & \cdots & \uno_{n_2} & 0 \\
\vdots & & & & \vdots \\
0 & \uno_{n_2} & \cdots & 0 & 0 \\
\uno_{n_1} & 0 & \cdots & 0 & 0
\end{array}\!\!}
\quad ; \quad
\gamma^-_{\Omega,9}=\inpar{\!\!\begin{array}{ccccc}
0 & 0 & \cdots & 0 & i\uno_{n_1}\\
0 & 0 & \cdots & i\uno_{n_2} & 0 \\
\vdots & & & & \vdots \\
0 & -i\uno_{n_2} & \cdots & 0 & 0 \\
-i\uno_{n_1} & 0 & \cdots & 0 & 0
\end{array}\!\!}  \  ,
\label{gomegap}
\eeq
for $\epsilon=1$ and $\epsilon=-1$ respectively.

Given the realization of the full orientifold action we can
proceed to find the Chan-Paton factors for the various massless
and tachyonic open states.
The gauge group for both $\epsilon=\pm 1$ turns out to be
\beq
G_9= \u(n_1) \times \cdots \times \u(n_L)  \  ,
\label{g99op}
\eeq
and similarly for $G_{\bar 9}$.
The charged tachyons in the $9\bar 9$ sector transform under
$G_9 \times G_{\bar 9}$ as
\beq
\sum_{k=1}^L \,
\incor{\inpar{\repf_k,\repfc_{\bar k}}+ {\rm c.c.} }  \ .
\label{tachyop}
\eeq
The charged massless fermions depend on the specific $\Z_N$.
Examples will be presented in section \ref{ssecexamples}.
Actually it suffices to consider $\epsilon=1$, otherwise we just exchange branes
and antibranes.

\subsection{Examples}
\label{ssecexamples}

In this section we present a more detailed account of closed and open states in 
some $\Z_N^*$ selected from those in (\ref{tab1}). Since the generic
gauge groups are already given in eqs.~(\ref{g99om}) and (\ref{g99op}), we will
only supply the relations between $n_k$
and $\bar n_k$ due to RR tadpole cancellation. To simplify
presentation we only display the massless charged fermions. We also provide the
closed massless and tachyonic states. In the untwisted sector, denoted $\theta^0$,
we only list the states besides the dilaton plus metric, and the antisymmetric
tensor, that are always present among respectively NSNS and RR massless states.
In the twisted $\theta^n$ sectors we list all states.
For $N$ even we will consider the $\Omega$ and $\op$ projections. 
The closed spectrum in both cases differ only in the $\theta^\nm$ sector.  

It is straightforward to work out many more examples. In table \ref{tabla}
we display the results for the remaining $\Z_6^*$ in (\ref{tab1}), but only with
$\op$ projection because then D9-branes must be necessarily included and antibranes can
be completely removed.

\subsubsection{${\mathbf \deq10},$  $\Z_2^*,$  $v=(0,0,0,1).$}

With $\Omega$ projection
this is the type 0B orientifold discussed at length in \cite{bs, Sagnotti, bg}.
The RR tadpoles cancel if $n_0=\bar n_0$, and $n_1=\bar n_1$. Clearly, we
are free to discard branes and antibranes altogether. However, we will keep
them because NSNS tadpoles cancel if $n_0+n_1=32$.
The closed spectrum in the twisted sector is:
\beq
%\theta^0 & : & \ \ \ \incor{ \r1 + \r{35}} \smim{(NSNS)} + \r{28} \, \smim{(RR)} \  ,
%\nonumber \\[0.2cm]
\theta \ \ \  :  \ \ \ \r1_{-1}  \, \smim{(NSNS)} + \r{28} \, \smim{(RR)} \  ,
\label{sz2c}
\eeq
where states are labelled by their $\so(8)$ representations and $\r1_{-1}$
denotes a tachyon of mass $-1$ ($\alpha^\prime$=1). The charged massless fermions are:
\beq
\r8_s \inpar{\repf_0,\repf_1} +\r8_c \inpar{\repf_0,\repf_{\bar 1}} +
\msm{k \leftrightarrow \bar k}  \ .
\label{sz2o}
\eeq
For $\epsilon=-1$ there are $\us$ rather than $\so$ groups but the fermions
transform in the same way.

With $\op$ projection we have instead
the type ${\rm 0B}^\prime$ orientifold first discussed in \cite{Sagnotti}.
The closed spectrum in the twisted sector is now:
\beq
\theta \ \ \  :  \ \ \ \incor{\r1 + \r{35}^+ } \smim{(RR)} \  .
\label{sz2cop}
\eeq
Comparing with (\ref{sz2c}) we see that the tachyon has disappeared and instead
of an antisymmetric tensor there is a 4-form with self-dual field strength,
denoted $\r{35}^+$.

The RR tadpole cancellation condition with the $\op$ projection is
$n_1=32+\bar n_1$. The charged massless fermions are:
\beq
\r8_s  \incor{\repa_1 + \reps_{\bar 1} + {\rm c.c.} }  +
\r8_c \incor{\inpar{\repf_1,\repf_{\bar 1}} + {\rm c.c.} }  \ .
\label{sz2oop}
\eeq
The irreducible $\tr R^6$ anomaly of the 4-form is precisely cancelled by these 
fermions \cite{Sagnotti, sw}.

\subsubsection{${\mathbf \deq8},$  $\Z_3^*,$  $v=(0,0,0,\frac23).$}

The closed sector states, identified by $\so(6)$ representations, are:
\beqa
\theta^0 & : & \ \ \ \r1 \,  \smim{(NSNS)} + \r1 \, \smim{(RR)}
+ \r4 \, \smim{(NSR)}  \  ,
\nonumber \\[0.2cm]
\theta + \theta^2 & : & \ \ \ 3  \inll{ \incor{\r1_{-\frac23} + \r1} \smim{(NSNS)} +
\incor{\r1 + \r{15}} \smim{(RR)}
+ \r4 \, \smim{(NSR)} }\  .
\label{sz3c}
\eeqa
With orthogonal projection ($\epsilon=1$),
the RR (and NSNS) tadpoles cancel if $n_0=\bar n_0$, and $n_1=16+\bar n_1$.
Notice that antibranes can be avoided but we will give the generic open spectrum. 
The charged massless fermions are:
\beq
\r4 \incor{ \repa_1 + \reps_{\bar 1} } + \bar{\r4} \inpar{\repf_1, \repf_{\bar 1}}
+ \inll{ \r4 \inpar{\repf_0,\repfc_1} +
\bar{\r4} \inpar{\repf_0,\repfc_{\bar 1}} + \msm{k \leftrightarrow \bar k} }  \ .
\label{sz3o}
\eeq
It is easy to check that irreducible $\tr F^5$ gauge anomalies are absent by virtue of
RR tadpole cancellation.

With symplectic projection, $G_{9}= {\rm USp}(n_0)\times \u(n_1)$ and 
$G_{\bar 9}={\rm USp}(\bar{n}_0)\times \u(\bar{n}_2)$, with even  
$\bar n_0=n_0$ and  $\bar n_1=16 + n_1$. Massless fermions are basically those
of the orthogonal case upon interchanging representations of
$G_{9}$ and $G_{\bar 9}$. The total anomaly factorizes appropriately.

\subsubsection{${\mathbf \deq6},$  $\Z_5^*,$  $v=(0,0,\frac15,\frac35).$}

The closed sector states are:
\beqa
\theta^0 & : & \ \ \ 2(\r1) \, \smim{(NSNS)} + 2(\r1) \, \smim{(RR)} +
2\incor{\r2_{\rm L} + \r2_{\rm R}} \smim{(NSR)}   \  ,
\nonumber \\[0.2cm]
\theta + \theta^4 & : & \ \ \ 5  \inll{
\incor{\r1_{-\frac25} + \r1} \smim{(NSNS)} + \incor{\r1 + \r3^+} \smim{(RR)}
+ 2(\r2_{\rm L}) \, \smim{(NSR)} } \  ,
\label{sz5c} \\[0.2cm]
\theta^2 + \theta^3 & : & \ \ \ 5  \inll{
\incor{\r1_{-\frac25} + \r1} \smim{(NSNS)} + \incor{\r1 + \r3^-} \smim{(RR)}
+ 2(\r2_{\rm R}) \, \smim{(NSR)} } \  ,
\nonumber
\eeqa
Here we label states by their $\so(4)$ representations and $\r3^+$ ($\r3^-$) denotes a self-dual
(antiself-dual) tensor. In $\su(2) \times \su(2)$ notation in which $\r2_{\rm L}=(\msm{\oh},0)$,
$\r3^+=(1,0)$.

When $\epsilon=1$ both RR and NSNS tadpoles are absent provided that
$n_0=\bar n_0$, $n_1=\bar n_1 + 8$, and $n_2=\bar n_2+8$.
It is possible to keep only D9-branes.
The generic charged massless fermions are:
\beqa
&\r2_{\rm L} & \!\!\! \incor{ \inpar{\repf_0,\repf_2} + \inpar{\repf_1,\repf_2}
+ \inpar{\repf_0,\repf_{\bar 1}} + \inpar{\repf_1,\repfc_{\bar 2}} +
\msm{k \leftrightarrow \bar k} } +
\nonumber \\[0.2cm]
& \r2_{\rm R} &  \!\!\! \incor{ \inpar{\repf_0,\repf_1} + \inpar{\repf_1,\repfc_2}
+ \inpar{\repf_0,\repf_{\bar 2}} + \inpar{\repf_1,\repf_{\bar 2}} +
\msm{k \leftrightarrow \bar k} } +
\label{sz5o} \\[0.2cm]
& \r2_{\rm L} & \!\!\!  \incor{\repa_1 + \reps_{\bar 1} +
\inpar{\repf_2,\repf_{\bar 2}} }  +
\,  \r2_{\rm R}  \incor{\repa_2 + \reps_{\bar 2} +
\inpar{\repf_1,\repf_{\bar 1}} } \ ,
\nonumber
\eeqa
plus the complex conjugates.
The irreducible $\tr R^4$ and $\tr F^4$ anomalies cancel.

With $\epsilon=-1$ the gauge groups change into  $G_{9}=
\us(n_0)\times\u(n_1)\times \u(n_2)$ and
$G_{\bar 9}=\us(\bar{n}_0)\times\u(\bar{n}_1)\times \u(\bar{n}_2)$, 
where $n_0$ and $\bar{n}_0$ must be even.
Tadpole cancellation requires $n_0=\bar{n}_0$, $\bar n_1=8+n_1$, and
$\bar n_2=8 + n_2$. Massless fermions follow from (\ref{sz5o})
upon interchanging representations of $G_{9}$ and $G_{\bar 9}$. 
The irreducible anomalies cancel as expected.

\subsubsection{${\mathbf \deq4},$  $\Z_6^*,$  $v=(0,\frac13,\frac13,\frac13).$}

We study first the $\Omega$ projection.
The closed sector spectrum is:
\beqa
\theta^0 & : & \ \ \ 9(\msm{0}) \smim{(NSNS)} + 10(\msm{0}) \smim{(RR)}   \  ,
\nonumber \\[0.2cm]
\theta + \theta^5 & : & \ \ \ 27 (\msm{0}) \smim{(RR)}\  ,
\nonumber \\[0.2cm]
\theta^2 + \theta^4 & : & \ \ \ 27 \inll{ \msm{0} \smim{(NSNS)} + \msm{0} \smim{(RR)} }\  ,
\label{sz6d4c} \\[0.2cm]
\theta^3 & : & \ \ \ \msm{0}_{-1} \, \smim{(NSNS)} + 10(\msm{0}) \smim{(RR)}   \  .
\nonumber
\eeqa
States are now labelled by helicity, $\msm{0}$ or $\msm{\pm \oh}$.

As in all $\Z_6^*$ examples, RR tadpoles vanish when $n_k=\bar n_k$, $k=0,\cdots,3$.
We also find that to cancel NSNS tadpoles, when $\epsilon=1$, $n_0+n_3=8$ and 
$n_1+n_2=12$. The charged massless fermions are:
\beqa
&  \!\!\!\!\! &
(\msm{-\oh})\inll{
3\incor{ \inpar{\repf_0,\repfc_{\bar 1}} +
\inpar{\repf_1,\repfc_{\bar 2}} + \inpar{\repf_2,\repf_{\bar 3}} } +
 \incor {\inpar{\repf_0,\repf_{\bar 3}} +
 \inpar{\repf_1,\repf_{\bar 2}}  + \inpar{\repfc_1, \repfc_{\bar 2}} } } +
%\nonumber \\[0.2cm]
\label{sz6d4o} \\[0.2cm]
&  \!\!\!\!\! & \!\!\ (\msm{\oh}) \inll{
3\incor{ \inpar{\repf_0,\repfc_1} + \inpar{\repf_1,\repfc_2} + \inpar{\repf_2,\repf_3} } +
\incor {\inpar{\repf_0,\repf_3} + \inpar{\repf_1,\repf_2}  + \inpar{\repfc_1, \repfc_2} } }
+ \msm{k \leftrightarrow \bar k}   \ .
\nonumber
\eeqa
Irreducible $\tr F^3$ gauge anomalies cancel.

Let us now the consider the model with $\op$ projection.
The closed $\theta^3$ sector includes instead a vector plus scalars. This is:
\beq
\theta^3 \ \ \  :  \ \ \  \incor{(\msm{\pm 1}) + 10(\msm{0})} \smim{(RR)}   \  .
\label{sz6d4cop}
\eeq
In the other twisted sectors tachyons are absent. Thus, this model has no closed tachyons.

Tadpoles cancel when $n_1=12+\bar n_1$, $n_2=8+\bar n_2$, and $n_3=12+\bar n_3$.
The massless charged fermions are:
\beqa
&(\msm{\oh}) & \!\!\! \inll{ 3\incor{ \inpar{\repf_1,\repfc_2} + \inpar{\repf_2,\repfc_3} }
+ \incor{ \inpar{\repf_1,\repf_3} + {\rm c.c.}} + \msm{k \leftrightarrow \bar k} } +
\nonumber \\[0.2cm]
& (\msm{\oh}) &  \!\!\! \inll{ 3\incor{ \repac_1 + \repa_3 + \repsc_{\bar1} + \reps_{\bar 3} } +
\incor{ \repa_2 + \repa_{\bar 2} + {\rm c.c.}} } +
\label{sz6d4oop} \\[0.2cm]
&(\msm{-\oh}) & \!\!\! \inll{ 3\incor{ \inpar{\repf_1,\repfc_{\bar 2}} +
 \inpar{\repf_2,\repfc_{\bar 3}} }
+ \incor{ \inpar{\repf_1,\repf_{\bar 3}} + {\rm c.c.}} + \msm{k \leftrightarrow \bar k} } +
\nonumber \\[0.2cm]
&(\msm{-\oh}) & \!\!\! \inll{ 3\incor{ \inpar{\repfc_1,\repfc_{\bar 1}} +
\inpar{\repf_3,\repf_{\bar 3}} }
+ \incor{ \inpar{\repf_2,\repf_{\bar 2}} + {\rm c.c.}} } \ .
\nonumber
\eeqa
The spectrum is free of gauge anomalies. For $\bar n_k=0$ it coincides with the results of
\cite{Angelantonj98, bfl} for the non-tachyonic ${\rm 0B}^\prime$
orientifold on $\T^6/\Z_3^{\rm susy}$.
This is to be expected since here the $\Z_6^*$ can be written as
$\Z_3^{\rm susy} \times \Z_2^*$, where $\Z_3^{\rm susy}$ has twist vector
$v=(0,\frac13,\frac13,-\frac23)$, and $\Z_2^*$ is $(-1)^{F_S}$.

\begin{table}[htb]
\renewcommand{\arraystretch}{1.25}
\footnotesize
\begin{center}
\begin{tabular}{|c||c|c|}
\hline
Sector & $\deq8$,  $v=(0,0,0,\frac13)$ & 
$\deq6$,  $v=(0,0,\frac13,\frac23)$ \\
\hline 
%$G_9$ & $\u(16) \times \u(16)$ &  $\u(8) \times \u(16) \times \u(8)$ \\ 
%\hline
$\theta^0$ & $\r1 + \r1$   & 
$4(\r1)   + 4(\r1) $    \\
$\theta + \theta^5$ &  
$3  \inll{ \r1_{-\frac13}  + \incor{\r1 + \r{15}} }$ & 
$9  \inll{\r1_{-\frac13} + \incor{\r1 + \r3^+}  }$ \\
$\theta^2 + \theta^4$ &  3  $\inll{ \incor{\r1_{-\frac23} + \r1} 
+ \incor{\r1 + \r{15}} }$  & 
$9  \inll{4(\r1)  + \incor{\r1 + \r3^-}  }$ \\
$\theta^3$ &  $\incor{\r1 + \r{15}}$ &  
$\incor{2(\r1) + \r3^+ + 3(\r3^-)}$  \\ 
\hline
 &  $\u(16) \times \u(16)$ &  $\u(8) \times \u(16) \times \u(8)$ \\ 
99 & $\r4 \incor{\inpar{\ov{\r{120}}, \r1}+ \inpar{\r1,\r{120}}}$ & 
$\r2_{\rm L} \incor{\inpar{\r8, \r1, \r8} + \inpar{\ov{\r8}, \r1, \ov{\r8}}
+ \inpar{\r1, \r{120}, \r1} + \inpar{\r1, \ov{\r{120}}, \r1} } + {\rm c.c.}$ \\ 
& & 
$\r2_{\rm R} \incor{\inpar{\r8, \ov{\r{16}}, \r1} + \inpar{\r1, \r{16}, \ov{\r8}}
+ \inpar{\r{28}, \r1, \r1} + \inpar{\r1, \r1, \r{28} } } + {\rm c.c.}$ \\ 
\hline
\end{tabular}
\end{center}
\caption{Closed tachyonic and massless states, gauge group, and massless fermions in $\Z_6^*$ orientifolds 
with $\op$ projection and no $\ov{\D 9}$.}
\label{tabla}
\end{table}

\section{Models with O9 and O5-planes}
\label{seco9o5}

To be concrete we study the case of ${\rm O}5_1$-planes that
occurs for the $\Z_N$ actions in (\ref{tab1}). In general these
have
\beq
v=(0,\frac{2k_1}{N}, \frac{2k_2+1}{N},  \frac{2k_3+1}{N})   \ ,
\label{v95N}
\eeq
with $k_i \in \Z$.

\subsection{Tadpole cancellation with O9 and O5-planes}
\label{sstco9o5}

It is convenient to organize tadpoles according to their
volume dependence.
We denote as $V_1$ the $V_P$ (or $V_P^{\rm NN}$)
invariant under $\theta^\nm$, and as $V_{23}$ the $V_W$ (or $V_W^{\rm
DD}$) also invariant under $\theta^\nm$. As
usual, the amplitudes $\cz_\ck(\uno, \uno)$, $\cz_{pp}(\uno)$ and
$\cz_p(\uno)$, $p=9, \bar 9$, produce an untwisted tadpole
proportional to $V_{10}$. Cancellation requires
\beq
\Tr \gamma_{0,9} - \Tr \gamma_{0,\bar 9} = 32 \, \epsilon_9
\  .
\label{tunt9}
\eeq
There are also tadpoles proportional to $V_D V_1/V_{23}$ in \deq4,
or to $V_D/V_{23}$ in \deq6, that arise from $\cz_\ck(\uno,
\theta^\nm)$, $\cz_{pp}(\uno)$ and $\cz_p(\theta^\nm)$, $p=5,
\bar 5$. These cancel provided
\beq
\Tr \gamma_{0,5} - \Tr \gamma_{0,\bar 5} = 32 \, \delta_5
\epsilon_5 \ ,
\label{tunt5}
\eeq
where we have used (\ref{deltadef}) and (\ref{epsilondef}).
We will shortly explain that tadpole cancellation
requires $\delta_p=-1$.

In the following we will assume the Gimon-Polchinski action
for $\Omega$, in particular, $\Omega^2=-1$ on 95 states \cite{gp}.
In our notation this means
\beq
\epsilon_9 = - \epsilon_5  \ .
\label{gpcon}
\eeq
This implies for instance that in absence of $\ov{\D 9}$-branes
necessarily $\epsilon_5=-1$, because  (\ref{tunt9}) demands $\epsilon_9=1$.

We now discuss twisted tadpoles. Neither the Klein nor the
M\"obius amplitudes contribute to tadpoles due to massless states
in sectors $\theta^m$ with $m$ odd. In this case there are only
cylinder tadpoles basically given by (\ref{tadrr95sum}). For $m$
odd we then have a cancellation condition
\beq
\inpar{\Tr \gamma_{m,9} - \Tr \gamma_{m,\bar 9}} +
\xi(mv) \,
\sqrt{\tilde\chi(\theta^m_{\rm DD})} \inpar{\Tr \, \gamma_{m,5,J}
- \Tr \gamma_{m,\bar 5,J}} = 0 \ ,
\label{tad95odd}
\eeq
with $J=1, \cdots, \tilde\chi(\theta^m_{\rm DD})$.

Let us now discuss tadpoles due to massless states in sectors
$\theta^m$ with $m=2\ell$ so that $T_\ck^{\rm RR}(n,\ell+k)$ and
$T_p^{\rm RR}(\ell+k)$, $n,\, k=0, \frac{N}2$, in principle do
contribute. To analyze these tadpoles it is convenient to
distinguish two cases depending on whether $\ell v_j$, for
transverse directions to the $\D 5_1$-branes, is a half-integer or
not.

\medskip
\noindent \underline{ $\ell v_j \notin \Z + \oh$, $j=2,3$.}
\medskip

\noindent
In this case the cylinder tadpoles add as shown in
(\ref{tadrr95sum}). The Klein tadpoles depend crucially on $v_1$.
If $\ell v_1 \in \Z + \oh$, Klein and cylinder tadpoles have a
different volume dependence. Furthermore there will be uncancelled
tadpoles, proportional to $V_D/V_1$, because
\beq
\sum_{n=0,\frac{N}2} T_\ck^{\rm RR}(n,\ell+k) \not= 0 \quad ;
\quad k=0, \nm \ .
\label{exk}
\eeq
This occurs in the supersymmetric $\Z_4$, $\Z_8$, $\Z_8^\prime$,
and $\Z_{12}^\prime$ in \deq4 \cite{afiv}. It also happens in the
non-supersymmetric $\Z_8$ with $v=\frac18(0,2,1,5)$ and $\Z_{12}$
with $v=\frac1{12}(0,2,1,5)$. As advocated in \cite{ru}, these problematic twisted
tadpoles can be cancelled by adding pairs of $\D 5_2$-$\ov{\D 5}_2$ and/or
$\D 5_3$-$\ov{\D 5}_3$ branes, besides the $\D 9$ and $\D 5_1$ required
by cancellation of untwisted tadpoles.

Consider next $\ell v_1 \notin \Z + \oh$. Then, all cylinder,
M\"obius and Klein tadpoles have the same volume dependence $V_D
V_P^{\rm NN}$. In the appendix the relevant M\"obius tadpoles
from 9 and 5-branes are  combined according to the sector.
Comparing the M\"obius result (\ref{tadrr95mob1}) with the cylinder tadpoles
(\ref{tadrr95sum}), with $m=2\ell$, we observe that the
same combination of $\Tr \,\gamma_{2\ell,9}$ and $\Tr \, \gamma_{2\ell,5,J}$
appears only if $\epsilon_9 =-\epsilon_5$. But this is precisely
the GP condition. Similarly, comparing (\ref{tadrr95mob2})
with the cylinder tadpoles we conclude that to have
the same combination of traces it must be that $\delta_9 = \delta_5$.

The Klein tadpoles $T_\ck^{\rm RR}(n,\ell+k)$,  $n,\, k=0, \frac{N}2$,
are directly given by (\ref{tadkbrrfull}) but they must be distributed among
the fixed points. In particular, the contribution to the fixed point at
the origin, denoted $T_{\ck,1}^{\rm RR}(\ell)$, turns out to be
\beq
T_{\ck,1}^{\rm RR}(\ell)
=   2^{D+I_P} \, V_D V_P^{\rm NN}
\sqrt{\frac{\tilde\chi(\theta^{2\ell}_{\rm NN})}{\tilde\chi(\theta^{2\ell}_{\rm DD})}}\,
\incor{\sqrt{\tilde\chi(\theta^{\ell+ \nm}_{\rm DD})} + \xi(2\ell v)
\sqrt{\tilde\chi(\theta^{\ell}_{\rm DD})} }^2 \ .
\label{tadrrk0}
\eeq
Simultaneous fixed points of $\theta^{\ell}$ and $\theta^{\ell+ \nm}$,
hence also of $\theta^{\nm}$, receive the same contribution as that of $J=1$.
Fixed points of $\theta^{\ell}$ alone get a share only
from  $T_\ck^{\rm RR}(0,\ell)$, this is given by the first term in (\ref{tadrrk0})
expanding the square. Similarly,  fixed points of $\theta^{\ell + \nm}$ alone get a share only
from  $T_\ck^{\rm RR}(0,\ell + \nm)$, given by the last term in (\ref{tadrrk0})
after expanding the square.  Fixed points of $\theta^{2\ell}$
only do not have Klein tadpoles.
Finally, when we compare the above Klein tadpole at the origin
with that due to the M\"obius amplitudes, c.f. (\ref{tadrr95mob0}),
we see that necessarily $\delta_9=-1$.

We are now ready to add the cylinder, Klein and M\"obius tadpoles.
The above results show that they add to a common prefactor times
a perfect square. Including antibranes we then find the cancellation
condition
\beqa
& &
\inpar{\Tr \gamma_{2\ell,9}  - \Tr \gamma_{2\ell,\bar 9}}
\, +  \, \xi(2\ell v) \, \sqrt{\tilde\chi(\theta^{2\ell}_{\rm DD})} \,
\inpar{ \Tr \gamma_{2\ell,5,J} - \Tr \gamma_{2\ell,\bar 5,J} } =
\nonumber \\[0.2cm]
& & \ \ \  \  \ \ \    2^{\oh(D+I_P)} \, \epsilon_9
\prod_{a=0}^3 c(\ell v_a)\,
\incor{\sqrt{\tilde\chi(\theta^{\ell+ \nm}_{\rm DD})} + \xi(2\ell v)
\sqrt{\tilde\chi(\theta^{\ell}_{\rm DD})} } \ ,
\label{tad95c1}
\eeqa
where $I_P$ is the dimension of the momentum sub-lattice invariant under
$\theta^{2\ell}$. This is valid when $J$ refers to a simultaneous fixed
point of $\theta^{\ell}$ and $\theta^{\ell+ \nm}$, such as the origin.
For fixed points of $\theta^{\ell}$ alone only
the first term in the right hand side of (\ref{tad95c1}) appears.
For fixed points of $\theta^{\ell + \nm}$ is only
the second term that appears. For fixed points of $\theta^{2\ell}$
alone the right hand side is zero.

When $N$ is a multiple of four we can consider $\ell=\frac{N}4$.
Then, the above results apply if $\frac{N}4 v_1 \notin \Z + \oh$.
In this case we can check that the quantity inside
brackets in (\ref{tad95c1}) vanishes identically. We then
find the tadpole cancellation condition
\beq
\inpar{\Tr \gamma_{\frac{N}2,9} - \Tr \gamma_{\frac{N}2,\bar 9}}
-4 \inpar{\Tr \, \gamma_{\frac{N}2,5,J} - \Tr \,
\gamma_{\frac{N}2,\bar 5,J}} = 0 \quad ; \quad J=1, \cdots, 16 \ .
\label{tad95N4}
\eeq
Examples are given by the $\Z_4$ in \deq6, supersymmetric or not.
Recall however that if $\frac{N}4 v_1 \in \Z + \oh$, the Klein and
cylinder tadpoles have a different volume dependence and moreover
there will be uncancelled tadpoles, proportional to $V_D/V_1$, as
in the supersymmetric $\Z_4$ in \deq4 \cite{afiv}.

\medskip
\noindent \underline{$\ell v_j \in \Z + \oh$, for $j=2$ and/or
$j=3$.}
\medskip

\noindent
In this case it happens that all RR tadpoles from $\cz_{95}(\theta^{2\ell})$,
$\cz_9(\theta^\ell)$, $\cz_5(\theta^{\ell+\nm})$, $\cz_\ck(\theta^\nm,\theta^\ell)$,
and  $\cz_\ck(\theta^\nm,\theta^{\ell + \nm})$, do vanish. Furthermore, the
tadpoles  $T_{99}^{\rm RR}(2\ell)$, $T_9^{\rm RR}(\ell+\nm)$ and $T_\ck^{\rm RR}(0,\ell+\nm)$,
all have the same volume dependence $ V_D V_P$. Adding all
contributions gives the cancellation condition
\beq
\Tr \gamma_{2\ell,9}  - \Tr \gamma_{2\ell,\bar 9}
= 2^{\oh(D+I_P)} \, \epsilon_9 \delta_9
\prod_{a=0}^3 c(\ell v_a + \nm v_a)\ \ ,
\label{tad95c2}
\eeq
where $I_P$ is the dimension of the momentum sub-lattice invariant under
$\theta^{2\ell}$.
To obtain this result we have assumed that $\tilde\chi(\theta^{2\ell}_{\rm DD}) =
\tilde\chi(\theta^{\ell+\frac{N}2}_{\rm DD})$ which does hold for the
crystallographic actions. Notice that for $\ell=\nm$ the above cancellation
condition reproduces (\ref{tunt9}).

The remaining tadpoles  $T_{55}^{\rm RR}(2\ell)$, $T_5^{\rm RR}(\ell)$ and $T_\ck^{\rm RR}(0,\ell)$
all have the same volume dependence $V_D V_P^{\rm NN}/V_W^{\rm DD}$. The way these pieces combine
depends on the dimension of the invariant subspace of $\theta^{2\ell}$ in the DD directions,
When  $\theta^{2\ell}$ leaves invariant
the whole sub-space transverse to the 5-branes the tadpole cancellation condition
turns out to be
\beq
\Tr \gamma_{2\ell,5}  - \Tr \gamma_{2\ell,\bar 5}
= 2^{\oh(D+I_P)} \, \epsilon_5
\prod_{a=0}^3 c(\ell v_a + \nm v_a)\ \ .
\label{tad95c3}
\eeq
As a check, observe that for $\ell=\nm$ we recover the condition
(\ref{tunt5}). The remaining possibility is that $\theta^{2\ell}$
leaves invariant only one of the transverse complex directions,
say $Y^3$. Factorization requires that $\theta^\ell$ leaves fixed
only the origin ($J=1$) in the $Y^2$ direction, and this is true
for crystallographic actions. We then find a cancellation
condition
\beq
\Tr \gamma_{2\ell,5,J}  - \Tr \gamma_{2\ell,\bar 5,J} =
2^{\oh(D+I_P)} \, \epsilon_5 \prod_{a=0}^3 c(\ell v_a + \nm v_a)\
\quad ; \quad J=1 \ .
\label{tad95c4}
\eeq
For other fixed points of $\theta^{2\ell}$ in the $Y^2$ direction there is an
analogous condition but with the right hand side equal to zero.

When all D5-branes are located at the origin, T-duality
imposes that $\Tr \gamma_{2\ell,9}= \Tr \gamma_{2\ell,5}$.  Thus,
$\epsilon_9 \delta_9 = \epsilon_5$ and the GP condition implies $\delta_9=-1$.
Repeating the argument with $\ell \to \ell + \nm$ shows that $\delta_5=-1$.

\subsection{Examples}
\label{ssex95}

The tadpole cancellation conditions that we have derived apply to
all $\Z_N$ orientifolds with action of the form (\ref{v95N}).
These include our non-supersymmetric as well as supersymmetric
orientifolds considered previously \cite{gp, gj, dp, Zwart, afiv,
ah}. Our results apply when the D5-branes are located at generic
fixed points of $\theta$ or some power. We will exemplify the
solutions of the tadpole conditions, and the corresponding
spectra, only for a few cases but we will consider moving the
D5-branes to the bulk or to other fixed points different from the
origin.

We will first analyze three $\Z_6$ orientifolds in which the
distinct possibilities for even twisted tadpoles arise.
We then study a $\Z_8$ orientifold in order to further illustrate
the importance of having tadpole cancellation conditions
that take into account all fixed points and not only the origin.

In general, $\gamma_{1,9}$ and $\gamma_{1,5,J}$ are of the form
(\ref{gonestarop}), and analogous for antibranes. 
For $\Omega$ we take the realization given in (\ref{gomegap}).
Recall that we are assuming the GP action $\epsilon_5=-\epsilon_9$.
We will explicitly discuss examples with $\epsilon_9=1$.
We have found that the opposite orientifold projection, $\epsilon_9=-1$, 
simply leads to the replacement of branes by antibranes and, whenever they 
appear, of ${\rm USp}$ by  ${\rm SO}$ groups.

\subsubsection{$\mathbf \deq6$, $v=(0,0,\frac16,\frac56)$}

For this $\Z_6$ action we take the torus lattice to be the product
of two SU(3) root lattices. The only fixed point of $\theta$ is
the origin denoted $X_1$. The fixed points of $\theta^2$ are of
the form $(w_i,w_j)$, where $w_0=(0,0)$, whereas $w_1$ and $w_2$
are the SU(3) weights. Thus, besides the origin $X_1$, $\theta^2$
has eight fixed points labelled $X_J$. Notice also that $\theta^3$
has sixteen fixed points. The chiral massless fields coming from the
closed sector are shown in table \ref{closedz6}.

\begin{table}[htb]
\renewcommand{\arraystretch}{1.25}
\footnotesize
\begin{center}
\begin{tabular}{|c|c|c|c|c|}
\hline
Twist &  $\theta^0$ & $\theta, \theta^5$ & $\theta^2, \theta^4$ & $\theta^3$ \\
\hline
$v=(0,0,\frac16,\frac56)$ & 
$-$ & 
$\r3^+$ + 6($\r2_{\rm L}$) &   
5($\r3^-$) + 16($\r2_{\rm R}$) &
10($\r2_{\rm L}$) \\
\hline
$v=(0,0,\frac16,\frac12)$ & 
$2(\r2_{\rm R})$ & 
4($\r3^+$) + 8($\r2_{\rm L}$) &   
2($\r3^+$) + 2($\r3^-$) + 4($\r2_{\rm L})$ + $2(\r2_{\rm R})$ &
8($\r2_{\rm R}$) \\
\hline
\end{tabular}
\end{center}
\caption{Chiral massless tensors and fermions in closed
sectors of \deq6 $\Z_6$ orientifolds.}
\label{closedz6}
\end{table}

The model-dependent twisted tadpole cancellation conditions are
obtained from (\ref{tad95odd}) for $m=1,3$, and from
(\ref{tad95c1}) with $m=2$, and distinguishing among the fixed
points of $\theta^2$. We find
\beqa
\inpar{\Tr \gamma_{1,9} - \Tr \gamma_{1,\bar 9}} - \inpar{\Tr \,
\gamma_{1,5,1} - \Tr \, \gamma_{1,\bar 5,1}} & = & 0 \  ,
\nonumber \\[0.2cm]
\inpar{\Tr \gamma_{2,9} - \Tr \gamma_{2,\bar 9}} +3 \inpar{\Tr \,
\gamma_{2,5,1} - \Tr \, \gamma_{2,\bar 5,1}} & = & -32\epsilon_9 \ ,
\label{z6a2} \\[0.2cm]
\inpar{\Tr \gamma_{2,9} - \Tr \gamma_{2,\bar 9}} + 3 \inpar{\Tr \,
\gamma_{2,5,J} - \Tr \, \gamma_{2,\bar 5,J}} & = & -8\epsilon_9
\quad ; \quad J=2, \cdots, 9 \ ,
\nonumber \\[0.2cm]
\inpar{\Tr \gamma_{3,9} - \Tr \gamma_{3,\bar 9}} -4 \inpar{\Tr \,
\gamma_{3,5,K} - \Tr \, \gamma_{3,\bar 5,K}} & = & 0 \quad ; \quad
K=1, \cdots, 16 \ .
\nonumber
\eeqa
For the supersymmetric \deq6, $\Z_6$ orientifold with
$v=(0,0,\frac16,-\frac16)$, there arise similar conditions as
first derived in \cite{gj,dp}. However, in previous works the
condition involving fixed points of $\theta^2$ other than the origin, is often missed. 
As stressed in \cite{abiu}, such conditions are needed to ensure
anomaly cancellation.

We fix $\epsilon_9=1$ and focus on solutions without antibranes. 
We then deduce that there must always be a number of D5-branes sitting at
the origin. It also follows that D5-branes can be placed at either
all of the $X_J$ or none of them. In the case without branes at
the $X_J$ we can further move $12k$ D5-branes to the bulk, say
$2k$ to a point $X$ and the remaining to the images of $X$ under
$\theta$. In this situation the generic gauge group turns out to
be
\beqa
G_9 & = & \u(n) \times \u(8-n) \times \u(8)  \  , \nonumber \\
G_{5_1} & = & \u(n-2k) \times \u(8-n-2k) \times \u(8-2k)  \  ,
\label{gsz6a} \\
G_{5_X} & = & {\rm USp}(2k)  \  ,
\nonumber 
\eeqa
where $2k \leq n \leq (8-2k)$. Notice that only $k=0,1,2$, are
allowed.

The massless fermions in each sector are given in table
\ref{ferz6a}. It is easy to check that the irreducible $\tr F^4$
anomaly cancels for each group factor. For this it is crucial that
the 95 fermions have left chirality as required by the GSO
projection explained in section \ref{sss95}. The gravitational
anomaly also cancels since eq. (\ref{ar4}) is verified when we take
into account the closed string spectrum shown in table
\ref{closedz6}.

\begin{table}[htb]
\renewcommand{\arraystretch}{1.25}
\begin{center}
\begin{tabular}{|c|c|}
\hline
Sector & massless fermions  \\
\hline $99$, $5_15_1$ & $\r2_{\rm L}$: $\inpar{\repf, \repf, \r1}+
\inpar{\repfc,\repfc, \r1} +
\inpar{\r1, \r1, \repa}+ \inpar{\r1, \r1, \repac}+ {\rm
c.c.}$\\[0.1cm]
 & $\r2_{\rm R}$: $\inpar{\repf, \repfc, \r1}+ \inpar{\repf, \r1,\repf}
 + \inpar{\r1, \repfc,\repfc} + {\rm c.c.}$\\
\hline $95_1$ & $\r2_{\rm L}$: $\inpar{\repf, \r1, \r1; \repfc,
\r1, \r1} + \inpar{\r1, \repf,\r1; \r1, \repfc, \r1} + \inpar{\r1,
\r1, \repf;
\r1, \r1, \repfc} + {\rm c.c.}$\\
\hline $5_X 5_X$ & $\r2_{\rm L}$: $2 \times \repa$
\quad ; \quad $\r2_{\rm R}$: $2 \times \reps$ \\
\hline $95_X$ & $\r2_{\rm L}$: $\inpar{\repf, \r1, \r1; \repf} +
\inpar{\r1, \repf,\r1; \repf} + \inpar{\r1, \r1, \repf;\repf} + {\rm c.c.}$\\
\hline
\end{tabular}
\end{center}
\caption{Massless fermions in open sectors of the $\Z_6$, $v=(0,0,\frac16,\frac56)$, orientifold.}
\label{ferz6a}
\end{table}

We can also move branes from the origin to $X_J$ fixed points.
There can be configurations with sixteen branes left at
$X_1$ and the remaining distributed among the $X_J$. For example,
there is a solution with
\beqa
G_9 & = & \u(n) \times \u(10-n) \times \u(6)  \  , \nonumber \\
G_{5_1} & = & \u(n-4) \times \u(6-n) \times \u(6)  \  ,
\label{gsz6a2}  \\
G_{5_J} & = & {\rm SU}(2)^4  \  .
\nonumber
\eeqa
Clearly only $n=4,5,6$, are allowed. In this solution there are
two D5-branes at each of the eight $X_J$, but the gauge group is
only ${\rm SU}(2)^4$ because these fixed points form doublets
under $\theta$. The massless fermions in the 99, $5_15_1$ and
$95_1$ sectors are the same, {\em mutatis mutandis}, as those in
table \ref{ferz6a}. In the $5_I5_I$ sector, compared to the
$5_X5_X$ in table \ref{ferz6a}, we have to be careful that now
there is a $\theta^2$ projection that removes the symmetric
representation and leaves the antisymmetric that is a singlet in
this case. Similarly, in the $95_I$
sector the $\theta^2$ projection only allows $\r2_{\rm L}$
massless fermions transforming as
\beq
\inpar{\r1, \r1, \repf ;\repf} + {\rm c.c.}
\label{95Iz6}
\eeq
for each SU(2). The full massless spectrum is anomaly-free as
expected.

There is another solution with two D5 at each $X_J$ in which
$G_{5_J}={\rm U}(1)^4$, but we skip further details.
It is also possible to move D5-branes from the origin to the
fixed points of $\theta^3$.

\subsubsection{$\mathbf \deq6$, $v=(0,0,\frac16,\frac12)$}

The torus lattice is an SU(3) times an SO(4) root lattice. 
The element $\theta$ has four fixed points of type
$(0,0)\otimes(\frac{a}2,\frac{b}2)$, with $a,b=0,1$. 
The element $\theta^2$ leaves the complex direction $Y^3$ fixed,
whereas in the direction $Y^2$ the fixed points are the
origin and the weights $w_1$ and $w_2$ of SU(3).
The closed massless chiral
states are displayed in table \ref{closedz6}.

The cancellation conditions for odd twisted tadpoles
follow from (\ref{tad95odd}) but in this model the $\theta^2$ 
tadpoles follow instead from (\ref{tad95c2}) and (\ref{tad95c4}).
Explicitly they read
\beqa
\inpar{\Tr \gamma_{1,9} - \Tr \gamma_{1,\bar 9}} + 2\inpar{\Tr \,
\gamma_{1,5,J} - \Tr \, \gamma_{1,\bar 5,J}} & = & 0 \quad ; \quad J=1, \cdots, 4\  ,
\nonumber \\[0.2cm]
\inpar{\Tr \gamma_{2,9} - \Tr \gamma_{2,\bar 9}} = \inpar{\Tr \,
\gamma_{2,5,1} - \Tr \, \gamma_{2,\bar 5,1}} & = & 16\epsilon_9 \ ,
\label{z6b2} \\[0.2cm]
\inpar{\Tr \,\gamma_{2,5,L} - \Tr \, \gamma_{2,\bar 5,L}} & = & 0
\quad ; \quad L=2, 3 \ .
\nonumber 
\eeqa
The condition involving $\gamma_{3,p}$ is automatically satisfied
with our choice (\ref{gonestarop}). 

We set $\epsilon_9=1$ and declare that $\ov{\D 9}$-branes
are absent. Besides, in general there is a net number of 32
D5-branes plus in principle some $\D 5$-$\ov{\D 5}$ pairs. In fact,
if we try to move $12k$ branes from the origin to the bulk these pairs will
be manifest. Observe that all fixed points of $\theta$ are equivalent
but we choose to refer to the origin. 
The solution of the tadpole cancellation conditions with
minimum number of $\D 5$-$\ov{\D 5}$ pairs at the origin leads to gauge groups 
\beqa
G_9 & = & \u(8) \times \u(8)   \  , \nonumber \\
G_{5_1} & = & \u(8-2k) \times \u(8-2k) \quad ; \quad G_{\ov{5}_1}  =  \u(2k)  \  ,
\label{gsz6b}  \\
G_{5_X} & = & {\rm USp}(2k)  \  .
\nonumber
\eeqa
This solution has the property that tachyons from $5_1\ov{5}_1$ strings are
eliminated by the orbifold projection. Massless fermions from
$\ov{5}_1\ov{5}_1$ and $9\ov{5}_1$ strings are projected out too.
The spectrum of surviving massless charged
fermions is presented in table \ref{ferz6b}. 
It is easy to prove that irreducible gauge and gravitational anomalies
cancel. 

\begin{table}[htb]
\renewcommand{\arraystretch}{1.25}
\begin{center}
\begin{tabular}{|c|c|}
\hline
Sector & massless fermions  \\
\hline $99$, $5_15_1$ & $\r2_{\rm L}$: $\inpar{\repf, \repfc}
+ {\rm c.c.}$\\[0.1cm]
 & $\r2_{\rm R}$: $\inpar{\repa, \r1}+ \inpar{\r1,\repa} + {\rm c.c.}$\\
\hline $95_1$ & $\r2_{\rm R}$: $\inpar{\repf, \r1; \repfc, \r1} + \inpar{\r1, \repf; \r1, \repfc} 
+ {\rm c.c.}$\\
\hline $5_1\ov{5}_1$ & $\r2_{\rm L}$: $\inpar{\repf, \r1; \repfc}
+ \inpar{\r1, \repf; \repfc} + {\rm c.c.}$\\[0.1cm]
 & $\r2_{\rm R}$: $\inpar{\repf, \r1; \repf}
+ \inpar{\r1, \repf; \repf} + {\rm c.c.}$\\
\hline $5_X 5_X$ & $\r2_{\rm L}$: $2 \times \repa$
\quad ; \quad $\r2_{\rm R}$: $2 \times \reps$ \\
\hline $95_X$ & $\r2_{\rm L}$: $\inpar{\repf, \r1; \repf} +
\inpar{\r1, \repf; \repf} + {\rm c.c.}$\\
\hline
\end{tabular}
\end{center}
\caption{Massless fermions in open sectors of the $\Z_6$, $v=(0,0,\frac16,\frac12)$, orientifold.}
\label{ferz6b}
\end{table}

Instead of moving D5-branes from the origin, or actually from any of the
$\theta$ fixed points, to the bulk, we can place them in groups of $6k$
at the two fixed points, $L=2,3$, of $\theta^2$ (in the $Y^2$ direction) 
that are exchanged by $\theta$.
These D5-branes give rise to the group
\beq
G_{5_L} =  \u(2k) \times {\rm USp}(2k)  \  .
\label{gz6b2}
\eeq
The USp factor appears because the D5-branes sit at fixed tori, since the
element $\theta^2$ leaves the $Y^3$ direction invariant.
The states from $\D9$, $\D5_1$ and $\ov{\D5}_1$ are still those
given in table \ref{ferz6b}. The massless fermions from the $\D5_L$
turn out to be:
\beq 
\begin{array}{ll}
 5_L 5_L & \r2_{\rm L}: \inpar{\reps, \r1}
+ \inpar{\repf; \repf} + {\rm c.c.}\\[0.1cm]
 & \r2_{\rm R}: \inpar{\repa; \r1}
+ \inpar{\repf, \repf} + {\rm c.c.}\\[0.1cm]
 9 5_L & \r2_{\rm R}: \inpar{\repf, \r1; \repfc, \r1}
+ \inpar{\r1,\repf; \repf, \r1} + {\rm c.c.}  \ .
\end{array}
\label{extraferz6b2}
\eeq
We have verified that the full content is anomaly-free.

Finally, there is a configuration with no $\ov{\D 5}$ and the 32
D5 equally distributed at the four fixed points of $\theta$. The gauge groups
in this case are
\beq
G_9= \u(12-2n) \times \u(4+2n) \quad ; \quad
G_{5_J} =  \u(n) \times \u(4-n) \quad ; \quad J=1, \cdots, 4  \  .
\label{gz6b3}
\eeq
Notice that the rank of the group arising from D5-branes is sixteen.
The representations for the massless fermions coincide with those given
in table \ref{ferz6b} for $99$, $5_15_1$, and $95_1$.

\subsubsection{$\mathbf \deq4$, $v=(0,\frac13,\frac12,\frac12)$}

This $\Z_6$ rotation can be realized on the product of
${\rm SU}(3) \times {\rm SO}(4) \times  {\rm SO}(4)$  root lattices.
Since the D5-branes wrap the first sub-torus, only the sixteen
fixed points $X_J$ of $\theta$ in the transverse complex directions
$(Y^2,Y^3)$ are relevant. 
The tadpole cancellation conditions turn out to be
\beqa
\inpar{\Tr \gamma_{m,9} - \Tr \gamma_{m,\bar 9}} + 4\inpar{\Tr \,
\gamma_{m,5,J} - \Tr \, \gamma_{m,\bar 5,J}} & = & 0 \quad ; \quad m=1,3
\quad ; \quad J=1, \cdots, 16\  ,
\nonumber \\[0.2cm]
\inpar{\Tr \gamma_{2,9} - \Tr \gamma_{2,\bar 9}} = \inpar{\Tr \,
\gamma_{2,5} - \Tr \, \gamma_{2,\bar 5}} & = & 16\epsilon_9 \ .
\label{z6c2} 
\eeqa
We consider $\epsilon_9=1$ and assume that there are only 32 D9-branes
present. For D5-branes we allow some number of $\D 5$-$\ov{\D 5}$ pairs. 

There are tadpole-free solutions with only 32 D5-branes distributed among
the fixed points of $\theta$. For example, we can locate two D5-branes at each
$X_J$, thereby obtaining a model with gauge groups $G_9=\u(10) \times \u(6)$ and
$G_5=\u(1)^{16}$.
Now, when we try to move D5-branes to the bulk some $\D 5$-$\ov{\D 5}$ pairs
are forced to remain at fixed points. To see this, we first place 32 D5-branes
at some specific $X_J$ and then displace $2k$ D5-branes, plus their $2k$
$\theta$-images, to the bulk that is fixed by $\theta^2$. We then find that
necessarily $4k$ $\ov{\D 5}$-branes must remain at $X_J$. The setup
with least number of $\ov{\D 5}$'s  has gauge group   
\beqa
G_9 & = & \u(8) \times \u(8)   \  , \nonumber \\
G_{5_J} & = & \u(8) \times \u(8) \quad ; \quad G_{\ov{5}_J}  =  \u(2k)  \  ,
\label{gsz6c}  \\
G_{5_X} & = & {\rm USp}(2k)  \  .
\nonumber
\eeqa
Tachyons from $5_J\ov{5}_J$ and massless fermions from
$\ov{5}_J\ov{5}_J$ strings are removed by the orbifold projection. 
The spectrum of massless charged fermions is presented in table \ref{ferz6c}. 
It is not difficult to check that $\tr F^3$ anomalies cancel. 

\begin{table}[htb]
\renewcommand{\arraystretch}{1.25}
\begin{center}
\begin{tabular}{|c|c|}
\hline
Sector & massless fermions  \\
\hline 
$99$, $5_J5_J$ &  $(\msm{\oh})$: $2\times\incor{ \inpar{\repf, \repfc}
+  \inpar{\repac, \r1}+ \inpar{\r1,\repa} }$\\[0.1cm]
\hline 
$95_J$ & $(\msm{\oh})$: $\inpar{\repfc, \r1; \repfc, \r1} + \inpar{\r1, \repf; \r1, \repf} $\\
\hline 
$9\ov{5}_J$ & $(\msm{-\oh})$: $\inpar{\repf, \r1; \repfc} + \inpar{\r1, \repfc; \repf} $\\
\hline 
$5_J\ov{5}_J$ & $(\msm{-\oh})$: $2\times\incor{\inpar{\repf, \r1; \repfc}
+ \inpar{\r1, \repfc; \repf} + \inpar{\repfc, \r1; \repfc}
+ \inpar{\r1, \repf; \repf} }$\\[0.1cm]
\hline 
$95_X$ & $(\msm{\oh})$: $\inpar{\repf, \r1; \repf} +
\inpar{\r1, \repfc; \repf} $\\
\hline
\end{tabular}
\end{center}
\caption{Massless fermions in open sectors of the $\Z_6$, $v=(0,\frac13,\frac12,\frac12)$, orientifold.}
\label{ferz6c}
\end{table}

\subsubsection{$\mathbf \deq6$, $v=(0,0,\frac18,\frac38)$}

This $\Z_8$ rotation allows a realization on a 4-dimensional hypercubic
lattice with orthonormal basis $e_i$. The action is $\theta e_i=
e_{i+1}$, $i=1,2,3$, and $\theta e_4=-e_1$. The fixed points
of $\theta$, denoted $X_1$ and $X_2$, are respectively the origin and the 
point $(\oh, \oh, \oh, \oh)$ at the center of the fundamental lattice cell.
The element $\theta^2$ has four fixed points, the two above plus $X_3$ and $X_4$
given by $(\oh, 0, \oh, 0)$ and $(0, \oh, 0, \oh)$. 
The element $\theta^4$ has sixteen fixed points, the four $X_J$
that are singlets under $\theta^2$, plus another
twelve points that form doublets of $\theta^2$.

The closed string spectrum can be found using the results of ref.~\cite{fh}
and implementing the $\Omega$ projection. We find that there is a net number
of three tensors $\r3^+$, but equal numbers of $\r2_{\rm L}$ and $\r2_{\rm R}$.
Thus, there is gravitational anomaly that must be compensated by open string
massless fermions.

The significant tadpole cancellation equations are obtained from our general
analysis. They read:
\beqa
\Tr \gamma_{1,9} + \sqrt{2}\Tr \gamma_{1,5,J} & = & 0 \quad ; \quad J=1,2 \  ,
\nonumber \\[0.2cm]
\Tr \gamma_{2,9} + 2\Tr \gamma_{2,5,J} & = & 16\sqrt{2} \quad ; \quad J=1,2 \ ,
\label{z8c} \\[0.2cm]
\Tr \gamma_{2,9} + 2\Tr \gamma_{2,5,J} & = & 0 \quad ; \quad J=3,4 \ ,
\nonumber \\[0.2cm]
\Tr \gamma_{3,9} - \sqrt{2}\Tr \gamma_{3,5,J} & = & 0 \quad ; \quad J=1,2 \  .
\nonumber 
\eeqa
Here we have already set $\epsilon_9=1$ and assumed that antibranes are absent.

The conditions involving the origin agree with general results in ref.~\cite{ah}
that however do not take other fixed points into account.
Now, neglecting cancellation conditions at other fixed points leads to the wrong
outcome that all 32 D5-branes could be placed at the origin. Indeed, it is simple
to prove that such configuration is not anomaly-free. In our approach this
is evident from the fact that the equations (\ref{z8c}) imply that if there are
D5-branes located at $X_1$ then necessarily there is the same number at $X_2$.
We can have a generic arrangement with $(16-2k)$ D5-branes at each $X_1$ and $X_2$,
together with $2k$ at each $X_3$ and $X_4$.

As an example, with D5-branes only at $X_1$ and $X_2$, 
the tadpole-free solution for the $\gamma_{m,p}$ matrices
yields gauge groups $G_9=\u(j)^2 \times \u(8-j)^2$ and 
$G_{5_1}=G_{5_2}=\u(8-j)\times \u(j)$. We refrain from displaying the full list
of massless fermions that is straightforward to derive using our prescriptions.
To make the main observations it is enough to consider the simplest case $j=0$.
In this case the massless fermions are just:
\beq 
\begin{array}{ll}
 99 & \r2_{\rm L}: \inpar{\r8, \r8} + {\rm c.c.}\\[0.1cm]
 & \r2_{\rm R}: \inpar{\r8,\ov{\r8}} + \inpar{\r1, \r{28}} + {\rm c.c.}\\[0.1cm]
 5_1 5_1, 5_2 5_2   & \r2_{\rm R}: \r{28} + {\rm c.c.}  \ .
\end{array}
\label{ferz8}
\eeq
Clearly, $N_R - N_L=84$, as needed to cancel gravitational anomalies.
The $\tr F^4$ anomalies vanish as well. Note that in general
$G_9= G_{5_1} \times G_{5_2}$ so that we might say that the brane configuration
is T-dual. However, as the $j=0$ example shows, the matter content is
not 95 symmetric.

Another simple solution has sixteen D5-branes at each $X_3$ and $X_4$.
Since these points are exchanged by $\theta$, there is only a gauge
factor $G_5$ of rank eight, in fact $G_5=U(8)$. We also find $G_9=U(8)\times U(8)$
and massless fermions 
\beq 
\begin{array}{ll}
 99 
 & \r2_{\rm R}: \inpar{\r{28},\r1} + \inpar{\r1, \r{28}} + {\rm c.c.}\\[0.1cm]
 55 & \r2_{\rm R}: \r{28} + {\rm c.c.}  \ .
\end{array}
\label{ferz8b}
\eeq
We can argue that this model is connected to that in eq.~(\ref{ferz8}) upon
T-duality in the compact directions so that $\D 9 \leftrightarrow \D 5$. 
Clearly, the group $G_{5_1} \times G_{5_2}$ and its matter content match the new
99 sector. Concerning the starting 99 sector, the idea is that we turn on Wilson
lines to separate branes in two stacks of sixteen each so that the bifundamentals
become massive. To reduce the rank and keep $\Z_8$ invariance these stacks
must be at fixed points of $\theta^2$.

\section{Models with O9-planes and O7 or O3-planes}
\label{seco9o73}

The prototypical examples having O7 and O3-planes are furnished by the
orientifolds with $\Z_4$ action given by the twist vectors
$v=(0,0,0,\oh)$ and $v=(0,\oh, \oh,\oh)$ respectively.  
These non-supersymmetric orientifolds have been discussed in
refs. \cite{bk, dm}. Here we will basically reproduce their results
using our formalism. In particular, in contrast to \cite{bk, dm},
in our language from the beginning it is clear that a generator $\theta$
that just reflects an odd number of complex coordinates is truly of order
four. This is required by modular invariance and it is consistent with the fact
that such $\theta$ has order four acting on world-sheet fermions.
In this section we mainly discuss the $\Z_4$ examples but the same 
procedure can be applied to more general cases, for
example when $Nv=(0,4k_1, 4k_2, 4k_3+2)$, or $Nv=(0,4k_1+2, 4k_2+2, 4k_3+2)$,
$k_i \in \Z$, and $N$ is multiple of four.

A distinguishing property of this class of models is that $\Omega^2=(-1)^{F_S}$,
when acting on $pp$ strings, $p=7,3$, as can be shown by looking how
$\Omega$ acts on states in the Ramond sector. In turn this property
implies that
\beq
 \gamma_{\Omega ,p} \inpar{\gamma_{\Omega ,p}^T}^{-1} = \epsilon_p 
\gamma_{N/2 ,p} \ ,
\label{omega2d37}
\eeq
where we have used that $\theta^{N/2}=(-1)^{F_S}$. It then follows that
$\epsilon_p^4=1$. On the other hand, acting on $9p$ states we have the
GP action $\Omega^2=e^{i\pi(9-p)/4}$.
 
To discuss tadpole cancellation conditions we will center on the $\T^{9-p}/\Z_4$
orientifolds, with $p=7$ or $p=3$, that have O9-planes together with ${\rm O} p$-planes. 
This is enough to illustrate the main features. To start we analyze
the divergences in the Klein-bottle amplitude. Tadpoles originate
from $\cz_\ck(\uno, \theta^m)$ and $\cz_\ck(\theta^\nm,\theta^m)$, for
$m=0,2$, they are proportional to $V_{10}$, whereas for $m=1,3$, they
depend on $V_D/V_W$. 
When $\Omega_P=\Omega$, the RR tadpoles actually vanish, i.e.
eq.~(\ref{krr1}) also holds in this situation.
Contrariwise, when $\Omega_P=\Omega^\prime$ 
the divergences from the untwisted and the $\frac{N}2$-twisted
sector combine to yield a non-zero RR tadpole and D9 plus $\D p$-branes must be added. 
This is the most interesting case and we will exclusively focus on it in the following.

Tadpoles due to cylinder amplitudes $\cz_{pp}$ are given by eq.~(\ref{tadrrpp}).
Concerning  $\cz_{p9}$, it corresponds to the amplitude of a sector twisted by $v_{p9}$ 
with entries $\oh$ in the DN directions. We can take $v_{p9}=\frac{N}4 v$ for $p=3,7$.
Then there is a modified spin structure $s_{0\oh}=-e^{-i\pi N S_v/4}$ that enters
in RR tadpoles and must be taken into account in the GSO projection
of $p9$ states.
It also happens that if $Y^j$ is DN and $mv_j \in \Z$, the RR tadpole from 
$\cz_{p9}(\theta^m)$ is zero. Notice that the DN (ND) coordinates for $p9$ ($9p$)
strings are the same as the DD for $pp$ strings.
If $mv_j \notin \Z$ for the DD coordinates,
all cylinder tadpoles have the same volume dependence and we have a result analogous
to (\ref{tadrr95sum}) with the replacement 
\beq 
\xi(mv) \to \hat\xi(mv)=- e^{-i\pi \frac{N}4 S_v} \! \prod_{\rm DD} s(mv_j) \ ,
\label{famsign73} 
\eeq 
because the number of DD directions is now odd and there is a different $v_{p9}$.
 
The M\"obius amplitudes for D7 and D3-branes are also similar to those for
D5-branes explained in the appendix. In fact, the M\"obius tadpoles can be read
off from the formulas in section \ref{ssms} upon obvious modifications.  

Putting together all necessary results leads to tadpole cancellation conditions
that can be organized according to volume dependence. The odd twisted tadpoles
only receive contributions from cylinder amplitudes and are proportional to $V_D$.
For instance, in the $\Z_4$ in \deq8 we find
\beqa
\Tr \gamma_{1,9} + 2 i\Tr \gamma_{1,7,J} & = & 0 \quad ; \quad J=1,\cdots, 4 \  ,
\nonumber \\[0.2cm]
\Tr \gamma_{3,9} - 2 i \Tr \gamma_{3,7,J} & = & 0 \ .
\label{oddd7} 
\eeqa
To simplify we are not including antibranes.

Tadpoles proportional to $V_{10}$ give the condition
\beq
\inpar{\Tr \gamma_{0,9}}^2 +  \inpar{\Tr \gamma_{2,9}}^2 - 
2^6 \epsilon_9(1-\delta_9) \Tr \gamma_{0,9} + 2^{12}=0 \ ,
\label{tcv10}
\eeq
where we have used (\ref{epsilondef}) valid for D9-branes.
The solution must be $\Tr \gamma_{2,9}=0$, $\delta_9=-1$ and
\beq
\Tr \gamma_{0,9} = 64 \epsilon_9  \  .
\label{soltcv10}
\eeq
Since antibranes are absent, necessarily $ \epsilon_9=1$ and there
are 64 D9-branes.

Cancellation of the remaining tadpoles proportional to $V_D/V_W$ 
involves the D7, or D3-branes, and requires instead 
\beq
\inpar{\Tr \gamma_{0,p}}^2 +  \inpar{\Tr \gamma_{2,p}}^2 + 
2^6 \incor{\Tr \gamma^{-1}_{\Omega 1,p} \gamma^T_{\Omega 1,p} - 
\Tr \gamma^{-1}_{\Omega 3,p} \gamma^T_{\Omega 3,p}  } + 2^{12}=0 \ .
\label{tcv8v3}
\eeq 
The solution now is $\Tr \gamma_{2,p}=0$ together with 
\beq
\gamma^{-1}_{\Omega 1,p} \gamma^T_{\Omega 1,p} = - \gamma_{0,p} 
\quad ; \quad
\gamma^{-1}_{\Omega 3,p} \gamma^T_{\Omega 3,p} = \gamma_{0,p}  \ ,
\label{soltcv8v3}
\eeq
so that $\Tr \gamma_{0,p} = 64$.

We will now present the matrices that fulfill the above conditions. 
Consider first the D9 sector. Since $\gamma^4_{1,9}=-\uno$,
and there are 64 D9-branes, we can realize $\theta$ as 
\beq
\gamma_{1,9} = 
{\rm diag}(\nu\uno_{16}, \nu^3 \uno_{16}, \bar\nu^3 \uno_{16}, \bar\nu\uno_{16})
\quad ; \quad
\nu=e^{i\pi/4}
\ .
\label{gonep73}
\eeq
Notice that this is eq.~(\ref{gonestarop}) with $n_j=16$. Clearly,
$\Tr \gamma_{1,9} = \Tr \gamma_{2,9}= 0$. 
The world sheet parity $\op$ is represented by $\gamma^+_{\Omega,9}$
given in eq.~(\ref{gomegap}).

In the $\D p$ sector we will locate all 64 branes at the origin so that
we can choose $\gamma_{1,p}= \gamma_{1,9}$. The orientifold action
is instead given by 
\beq
\gamma_{\Omega,p}=\inpar{\!\!\begin{array}{cccc}
0 & 0 & 0 & \nu\uno_{16}\\
0 & 0 & \bar\nu\uno_{16} & 0 \\
0 & \nu\uno_{16} & 0 & 0 \\
\bar\nu\uno_{16} & 0 & 0 & 0
\end{array}\!\!} \ .
\label{gomegap73}
\eeq
Therefore, $\gamma_{\Omega m ,p} =  \gamma_{m ,p} \gamma_{\Omega ,p}$
satisfies the conditions (\ref{soltcv8v3}). Moreover,
\beq
 \gamma_{\Omega ,p} \inpar{\gamma_{\Omega ,p}^T}^{-1} = 
\gamma_{2 ,p} \ .
\label{checkomega2d37}
\eeq
Comparing with the general property (\ref{omega2d37}) we see that $\epsilon_p=1$.
Besides, notice that $\gamma^4_{1,p}=-\uno$.

Let us now study the massless open spectrum. The gauge groups are easy to
determine, we find $G_9 = G_p= \u(16) \times \u(16)$. 
In the \deq4 orientifold with D9 and D3-branes, fermions are labelled by their chirality
$\pm \ts{\frac12}$. To be concrete we look at this case and later state the results
for D7-branes. Using the general rules explained in section \ref{sss99} we
readily find that the 99 massless fermions are 
\beq
4(\msm{\oh})\incor{ \inpar{\r{120}, \r1}
+  \inpar{\r1,\r{120}}+ \inpar{\ov{\r{16}},\ov{\r{16}}} } \ .
\label{99ferd73}
\eeq
For 33 massless fermions we have to be careful that the action of $\Om$
on the Ramond sector includes a factor of $i$, that implies $\Om^2=(-1)^{F_S}$
as mentioned before. The $\Z_4$ and $\Om$ projection on 33 fermions of positive
chirality require a fermionic Chan-Paton factor satisfying   
\beq
\lambda_F = i
\gamma_{1,3}\lambda_F\gamma_{1,3}^{-1} \quad ; \quad
\lambda_F = -i\gamma_{\Omega,3} \lambda_F^T
\gamma_{\Omega,3}^{-1} \ .
\label{cp33}
\eeq
Then, the massless 33 fermions are also given by (\ref{99ferd73}).

It remains to analyze the 93 and 39 states. From the mixed cylinder amplitude we deduce
that in the NS sector there are neither tachyonic nor massless scalars.
In the Ramond sector there is only one massless fermion of negative chirality  
in 93 and one of positive chirality in 39. 
The $\Z_N$ projection on the 39 Chan-Paton factor is $\lambda= \gamma_{1,3}\lambda \gamma_{1,9}^{-1}$. 
We also have to ensure that $\Om^2$ is realized properly. In general, for
$p9$ states it must be that
\beq
\lambda = e^{i\pi(p-9)/4} \gamma_{\Omega ,p} \inpar{\gamma_{\Omega ,p}^T}^{-1} \lambda \
\gamma_{\Omega ,9}^T \gamma_{\Omega ,9}^{-1}   \  . 
\label{omega2cp}
\eeq
For  39 states this gives a non-trivial constraint
$\lambda = i \epsilon_3 \epsilon_9  \gamma_{N/2 ,3} \lambda$.
The upshot is that in the $\Z_4$ orientifold with D3-branes we find mixed 
massless fermions
\beq
(\msm{\oh})\incor{ \inpar{\r{16},\r1;\r{16}, \r1}
+ \inpar{\r1,\r{16}; \r1, \r{16}} } \ .
\label{93ferd73}
\eeq
The overall massless spectrum is anomaly-free as first observed in \cite{bk, dm}.

Other configurations are viable. For example, $2k$ D3-branes plus their images
can move to the bulk while $(64-4k)$ remain at the origin. The bulk branes
give rise to an unitary group, in fact the full $G_3$ is 
$\u(16-k) \times \u(16-k) \times \u(k)$ \cite{dm}.

The $\T^2/\Z_4$ orientifold is treated in the same fashion. Recall that
fermions in \deq8 are either $\r4$ or  $\ov{\r4}$ of SO(6). 
With all D7-branes at the origin, the 77,
as well as the 99, massless fermions are a  $\ov{\r4}$ transforming under
$\u(16)\times \u(16)$ as given in eq.~(\ref{99ferd73}).
From strings between D9 and D7-branes there are tachyons.
In the 97 sector there is also a  massless fermion $\ov{\r4}$ transforming 
as shown in eq.~(\ref{93ferd73}). The full spectrum has no $\tr F^5$ anomalies.
Different brane arrangements are feasible. For instance, with equal number
of D7-branes on top of the four $\theta$-fixed points, 
$G_7=\incor{\u(4)\times \u(4)}^4$, whereas with all branes in the bulk,
$G_7=\u(16)$. Details of the spectrum can be worked out applying the rules
explained above.

To finish we will briefly comment on states from closed strings. Since
these orientifolds include the element $(-1)^{F_S}$ there are only
bosons in the spectrum. The $T^6/\Z_4$ model has the interesting feature that
closed tachyons are absent altogether because
the $\op$ projection eliminates the tachyon in the $\theta^2$ sector
and only massive scalars appear in other twisted sectors \cite{bk}. 
In the $T^2/\Z_4$ case the $\theta^2$ tachyon is projected out too, but
there are tachyons in other twisted sectors.

\section{Final Comments}
\label{sfin}

The principal motivation behind this work was to show that toroidal
orientifolds of the non-supersymmetric type 0 strings can be treated
on the same footing as type II orientifolds in which the orbifold point
group explicitly breaks supersymmetry. This formulation has the advantage
of allowing a transparent description of the resulting $D\leq 10$ theories
using the same language employed in the study of supersymmetric orientifolds.

Modding by a non-supersymmetric $\Z_N$ introduces subtleties that
must be handled carefully. In some cases with $N$ even, modular invariance
demands that the order $N$ be actually twice that of the geometrical
rotation. In fact, in this situation the world-sheet fermions truly feel
a $\Z_N$ action. Hence, there are really $N$ twisted sectors to be considered.
Existence of an $\nm$-th twisted sector in these cases explains typical
features such as the appearance of a model-independent tachyon, the doubling
of RR forms, and the cancellation of crosscap RR tadpoles when the $\Omega$
projection is used. Alternatively, with $\op$ projection this tachyon is projected
out and crosscap RR tadpoles remain. Furthermore, the non-supersymmetric actions
require modified spin structures in the loop amplitudes that then show up in
tadpole cancellation conditions and in the GSO projections that determine the
string states. We have striven to present clear prescriptions that can be easily
applied in any example. 

We have derived tadpole cancellation conditions that are also valid in the
supersymmetric orientifolds and take into account the location of $\D p$-branes
in the transverse DD directions. In models with D9 and D5-branes we obtained 
general expressions depending on the fixed set structure of the geometrical
rotation. To our knowledge such detailed analysis had only been carried out
before in specific examples \cite{gp, afiv}. This is important because correct
twisted tadpole formulas at all fixed sets are necessary to find solutions
leading to anomaly-free spectra as it should.
We have illustrated this issue with a $\Z_8$ orientifold in which tadpole conditions 
at the origin alone cannot imply anomaly cancellation.

Although we have not explored this possibility, 
the findings in this paper could be used to build semi-realistic examples.
Upon T-duality, the models with D9 and D5-branes will include
D3 and D7-branes as in the type of bottom-up constructions
studied in refs.~\cite{aiqu, aa}.
Our results allow to find systematically the gauge theories living on
D-branes at non-supersymmetric $\Z_N$ singularities. 

Another application is in the
construction of \deq4 orientifolds with flux \cite{grana}. In fact, we can already proffer
a simple example. Consider, the $\Z_4$ orientifold with 64 D3-branes. In
principle we can completely cancel O3-plane tadpoles by switching on fluxes 
of the RR and NSNS 3-forms with $N_{\rm flux}=64$. Thus, the D3-branes disappear
and we are left just with D9-branes having $G_9=\u(16) \times \u(16)$ and
massless fermions given in (\ref{99ferd73}). By itself this content is 
anomalous but the cubic anomaly can be precisely cancelled by the
flux contribution \cite{Uranga, cu}.

%\vspace*{1cm}
\newpage

{\bf \large Acknowledgments}

We are grateful to G. Aldazabal, A. Hern\'andez, L. Ib\'a\~nez and A. Uranga for 
useful remarks. A.F. thanks the Max-Planck-Institut f\"ur Gravitationsphysik
for hospitality at several stages of this work. 
J.A.L. thanks the Escuela de F\'{\i}sica, Universidad Central de Venezuela,
for financial aid, and the Departamento de F\'{\i}sica Te\'orica, Universidad 
Aut\'onoma de Madrid, for hospitality and support while completing this paper

\vspace*{1cm}

%\section{Appendix: Loop amplitudes}
\section*{A  Appendix: Loop amplitudes and tadpoles}
\label{appA}
%\addcontentsline{toc}{section}{\hspace{13pt} Appendix A: Loop amplitudes }
\setcounter{equation}{0}
\renewcommand{\theequation}{A.\arabic{equation}}

In this appendix we study the 1-loop amplitudes needed to compute
tadpoles and the spectrum of states. We will proceed as in the supersymmetric
case and will closely follow \cite{afiv}, limiting ourselves to
describing the relevant changes particular to the non-supersymmetric setup.

Schematically, after going to the tree-level channel, the one-loop
amplitudes give rise to divergences of the form
\beq
(T^{\rm NSNS} - T^{\rm RR}) \int_0^\infty \, d\ell  \ ,
\label{treetad}
\eeq
where, say $T^{\rm RR}$ receives contributions  $T_\ck^{\rm RR}$,
$T_{pq}^{\rm RR}$ and  $T_p^{\rm RR}$ from the
amplitudes $\ck$, $\cc_{pq}$ and $\cam_p$.
These divergences are due to the exchange of massless scalars in the
tree-level closed string channels, NSNS or RR according to the
superscripts.
In absence of supersymmetry, the tadpoles $T^{\rm NSNS}$
and $T^{\rm RR}$ are not necessarily equal and only
$T^{\rm RR}$ must cancel by consistency \cite{pol}.
Below we discuss the
Klein bottle , cylinder and M\"obius strip amplitudes.
Once the amplitudes are written explicitly it is straightforward to extract
the RR tadpoles. The main results are collected and organized into
tadpole cancellation conditions presented according to the
existing ${\rm O}p$-planes in sections \ref{seco9}, \ref{seco9o5} and \ref{seco9o73}.
These conditions are valid in the supersymmetric case as well.

\subsection*{A.1 {\large{\bf Klein bottle amplitude}}}
%\subsection{Klein bottle amplitude}
\label{sskb}

Being a closed string amplitude, $\ck$ includes a sum over
twisted sectors and an orbifold projection. In general,
\beq
\ck= \frac{V_D}{2N} \sum_{n,m=0}^N \int_0^\infty  \frac{dt}{t} \,
(4\pi^2 \a^\prime t)^{-D/2} \, \cz_\ck(\theta^n, \theta^m)  \  ,
\label{ampk}
\eeq
where
\beq
\cz_\ck(\theta^n, \theta^m) = \Tr \inpar{ P_{\rm GSO}\,  \Omega_P
\, \theta^m \, e^{-2\pi t[L_0 + \bar L_0]} } \  .
\label{tracek}
\eeq
The Virasoro operators are those of the $\theta^n$-twisted sector.
This trace does not contain the contribution of the non-compact
momenta that already appears in (\ref{ampk}). In particular,
$V_D$ is the regularized volume of the non-compact space-time.
The GSO projection will be explained shortly.
Since  $\theta^n \rightarrow \theta^{N-n}$ under $\Omega_P$,
only the untwisted sector ($n=0$) and the $\theta^\nm$ sector,
when $N$ is even, enter in the trace.
The integral in (\ref{ampk}) diverges as $ t \to 0$, or
equivalently as $\ell=1/4t \to \infty$, as indicated in (\ref{treetad}).
Given the $\Omega_P$ insertion, these divergences are created by massless
states in the sector twisted by $\theta^{2m}$.

The bosonic contribution to $\cz_\ck(\theta^n, \theta^m)$ contains
an standard oscillator piece and for $n\not=0$, a zero mode factor
$\tilde\chi(\theta^n, \theta^m)$  equal to the number of points fixed
simultaneously by $\theta^n$ and $\theta^m$, in the sub-space where
$\theta^n$ acts non-trivially. There is also a sum over quantized
momenta and windings whenever  $\theta^n$ and $\Omega\theta^m$
leave simultaneously invariant some sub-lattices of $\Lambda^*$ and
$\Lambda$.
The invariant momentum sub-lattice is denoted $\Lambda_I^*$,
its dimension $I_P$  and its volume $V_P$, whereas the invariant
winding sub-lattice is denoted $\Lambda_I$, its dimension $I_W$
and its volume $V_W$
(the dependence on $n$ and $m$ is understood and will be dropped for
simplicity).
%The volume of the invariant momentum sub-lattice is denoted $V_P$,
%whereas that of the invariant winding sub-lattice is denoted $V_W$
For instance, in $\Z_N^*$ examples, $\Lambda_I=\{ 0 \}$, and moreover,
either $\Lambda_I^*=\Lambda^*$ so that $V_DV_P$ is a 10-dimensional
$V_{10}$, or $\Lambda_I^*=\{ 0 \}$ and $V_P=1$ by definition.

The fermionic contribution to (\ref{tracek}) can be written in terms
of $\vartheta$ functions but in the non-supersymmetric case
we have to carefully take into account that in the $\theta^n$ twisted
sector the spin-structure coefficients $s_{\a\b}$ depend on $n$.
In the untwisted sector the $s_{\a\b}(0)$ are the usual ones (for type IIB):
\beq
s_{00}(0)=-s_{0\oh}(0)=-s_{\oh 0}(0)=s_{\oh\oh}(0)=1 \ ,
\label{ssus}
\eeq
for both left and right movers. This corresponds to the GSO
projection
\beq
P_{\rm GSO} = \frac14 \inpar{1+(-1)^{f_L}}\inpar{1+(-1)^{f_R}} \  ,
\label{pgso}
\eeq
where $f_L$, $f_R$, are world-sheet fermion numbers.
We conclude that in the untwisted sector inserting $\Omega$ or
$\Omega(-1)^{f_R}$ in the trace gives the same result because
$\Omega$ alone forces $f_L=f_R$ in the GSO projection and
then multiplying by $(-1)^{f_R}$ has no effect. In fact, for both
$\Omega$ and $\op$ we have
\beq
\cz_\ck^{fer}(\uno,\theta^m) = \sum_{\a,\b=0,\oh} s_{\a\b}(0) \
\prod_{a=0}^3 \frac{\tilde\vartheta\left[\a \atop {\b +2mv_a} \right]
}{\tilde \eta} \ .
\label{tkferu}
\eeq
The tilde on $\tilde\vartheta$ and $\tilde\eta$ means that these functions
have argument $e^{-4\pi t}$.

The $s_{\a\b}(n)$ are the spin-structure coefficients that also appear
in the torus amplitude. This is needed to have a consistent
$\ts{\oh}(1+\Omega_P)$ projection. Now, it can be shown \cite{ah, ft}
that modular invariance of the torus amplitude requires
\beq
s_{00}(n)=-s_{\oh 0}(n)=1 \quad ; \quad
s_{0\oh}(n)=-s_{\oh\oh}(n)=- e^{-i\pi n S_v} \ .
\label{ssts}
\eeq
Incidentally, notice that $s_{0\oh}(0)=s_{0\oh}(N)$ implies (\ref{modinv}).

Returning to $\cz_\ck$, for $N$ odd only the untwisted sector
enters and with (\ref{tkferu}) we are finished. For $N$ even we
still need to analyze the $\theta^\nm$ sector. With
$\Omega=\Omega_P$ the fermionic piece of the amplitude is just
\beq
\cz_\ck^{fer}(\theta^\nm,\theta^m) = \sum_{\a,\b=0,\oh}
s_{\a\b}(\nm) \ \prod_{a=0}^3 \frac{\tilde\vartheta\left[{\a
+ \frac{N}2 v_a}
 \atop {\b +2mv_a} \right] }{\tilde \eta} \ .
\label{tkfert} 
\eeq 
Putting all pieces together we find that for $\Omega_P=\Omega$, the RR tadpole, conveniently
normalized, is given by 
\beq T_\ck^{\rm RR}(n,m) = e^{-i\pi n S_v} 2^{D+I_P+I_W} \frac{V_D V_P}{V_W} \,\,
\tilde\chi(\theta^n, \theta^m) \!\!\!\! 
\prod_{nv_j \in \Z  +  \frac12} \!\!\!\! c(2mv_j + \msm{\oh}) \!\!
\prod_{{nv_j \in \Z} \atop {2mv_j\notin\Z}}  \! \! \! 2 \inmod{\sin{\inpar{2\pi mv_j}}}  \  , 
\label{tadkbrrfull}
\eeq 
where we have defined 
\beq c(x)=\frac{\inmod{\cos \pi x}}{\cos \pi x} \quad ; \quad c(n+\oh) \equiv 0 \ .
\label{signcos} 
\eeq 
When $nv_j \in \Z + \oh$ and $2mv_j \in \Z$, this Klein tadpole does vanish. For instance,
$T_\ck^{\rm RR}(\frac{N}2,\frac{N}2)=0$ in the $\Z_{even}$ examples with O5-planes.

In some cases, in particular in the $\Z^*_{even}$ and the $\Z_4$
in $D=8,4$ in (\ref{tab1}), $\frac{N}2 v$ happens to be an
$\so(8)$ weight in the vector class and indeed, $\theta^\nm$ is
equivalent to $(-1)^{F_S}$. Then, closed states in the
$\theta^\nm$ sector are rather characterized by a weight
$r^\prime=r+\frac{N}2 v$. Moreover, in these examples,
$s_{0\oh}(\frac{N}2)=s_{\oh \oh}(\frac{N}2)=1$. Hence, the sum
over spin structure demands $\sum_a r^\prime_a= {\rm even}$ and
the GSO projection effectively changes to
\beq
P^-_{\rm GSO} = \frac14 \inpar{1-(-1)^{f_L}}\inpar{1-(-1)^{f_R}} \  .
\label{pgsominus}
\eeq
Therefore, the difference between inserting $\Omega$ and $\Omega
(-1)^{f_R}$ is an overall minus sign, i.e.
\beq
\cz_\ck^{\op}(\theta^\nm, \theta^m) = -
\cz_\ck^{\Omega}(\theta^\nm, \theta^m)  \  .
\label{chsign}
\eeq
We conclude that when $\Omega_P=\Omega^\prime$, the tadpole
$T_\ck^{\rm RR}(\frac{N}2,m)$ picks an extra minus sign. Clearly,
this minus sign affects the type of divergences. We will see that
in these examples with $\Omega$ projection the RR tadpoles due to
$\cz_\ck(\uno, \theta^m)$ and $\cz_\ck(\theta^\nm, \theta^m)$
cancel each other but there are left-over NSNS tadpoles. With
$\op$ projection the opposite happens. Furthermore, the overall
minus sign in $\cz_\ck^{\op}(\theta^\nm, \theta^m)$ helps to get
rid of tachyons. When  $\frac{N}2 v$ is an $\so(8)$ weight in the
vector class, it is easy to show that the mass formula and the GSO
projection always allow a tachyon in the $\theta^\nm$ sector,
namely the state with $r^\prime=0$. If $\Omega_P=\Omega$, the
tachyon survives the orientifold projection but it is projected
out for $\Omega_P=\op$. The reason is precisely the relative minus
sign between the Klein bottle trace and the corresponding
torus trace \cite{Sagnotti, as}.

When $\frac{N}2 v$ is such that $\theta^\nm$ is not equivalent
to $(-1)^{F_S}$ we cannot conclude that there is a change of GSO
projection in this sector. Thus, the Klein amplitude is the same
for both $\Omega$ and $\op$ as in supersymmetric orientifolds.

\subsection*{A.2 {\large{\bf Cylinder amplitudes}}}
%\subsection{Cylinder amplitudes}
\label{sscyl}

Typically, the Klein bottle amplitude has divergences proportional to
$V_{10}$ or $V_D$ that can be cancelled by adding D9-branes.
For the $\Z_N^*$ these are the only type of divergences.
For other actions lower dimensional branes are also needed.
In particular, all $\Z_6$ and the \deq6 $\Z_4$ in (\ref{tab1}),
have $\D 5_1$-branes.  For the $\Z_4$ in \deq8 and \deq4 there are respectively
$\D7_3$ and D3-branes.

Open strings between branes lead to to cylinder amplitudes given by
\beq
\cc_{pq}= \frac{V_D}{2N} \sum_{m=0}^N \int_0^\infty  \frac{dt}{t} \,
(8\pi^2 \a^\prime t)^{-D/2} \, \cz_{pq}(\theta^m)  \  ,
\label{ampmobius}
\eeq
where
\beq
\cz_{pq}(\theta^m) =  \Tr \inpar{P_{\rm GSO} \, \theta^m \, e^{-2\pi t L_0} } \  .
\label{tracecyl}
\eeq
The trace is over open string states with $\D p$ and $\D q$
branes at the endpoints.
The integral has divergences of type (\ref{treetad}) due
to massless states in the sector twisted by $\theta^m$.

Consider first the $\cz_{pp}$ cylinders. In this case the boundary
conditions are either Neumann-Neumann (NN) for longitudinal
or Dirichlet-Dirichlet (DD) for transverse directions. The fermionic
contribution is then just
\beq
\cz_{pp}^{fer}(\theta^m) = \sum_{\a,\b=0,\oh} s_{\a\b}(0) \
\prod_{a=0}^3 \frac{\vartheta\left[\a \atop {\b + mv_a} \right] }{\eta} \ ,
\label{tcylfer}
\eeq
with $s_{\a\b}(0)$ given in (\ref{ssus}). The way the Chan-Paton degrees of
freedom enter in the trace depends on the boundary conditions. For instance,
for $p=9$ all directions are NN and there is factor $\inpar{\Tr \gamma_{m,9}}^2$.
To study other $\D p$-branes we denote by $\theta^m_{\rm NN}$ and
$\theta^m_{\rm DD}$ the restriction of $\theta^m$ to the NN and DD directions
respectively. The $\D p$-branes are located at some points in the transverse
directions with DD boundary conditions. If these points are not fixed under
$\theta^m_{\rm DD}$, the trace vanishes. To obtain a non-zero trace there must be
$\D p$-branes sitting at fixed points of $\theta^m_{\rm DD}$. The number of such
points will be denoted $\tilde\chi(\theta^m_{\rm DD})$. Then , $\cz_{pp}(\theta^m)$
will have a factor
\beq
\sum_{J=1}^{\tilde\chi(\theta^m_{\rm DD})}  \inpar{\Tr \gamma_{m,p,J}}^2  \  .
\label{fixtr}
\eeq
Recall that in general $\tilde\chi(\theta^m)$ is given by
\beq
\tilde\chi(\theta^m) = \prod_{mv_j\notin
\Z} \! 4 \sin^2 \pi mv_j \ .
\label{chitil}
\eeq
The boundary conditions also determine the type of lattice sums that can appear in
the bosonic part of $\cz_{pp}(\theta^m)$. In the NN directions there will be
a sum over quantized momenta if $\Lambda^*$ has a sublattice, of volume
$V_P^{\rm NN}$, invariant under $\theta^m_{\rm NN}$. In the DD directions there
can instead be windings. The volume of the $\Lambda$ sub-lattice invariant
under $\theta^m_{\rm DD}$ will be denoted $V_W^{\rm DD}$.

{}From the $\cc_{pp}$ amplitude we find the RR tadpole:
\beq
T_{pp}^{\rm RR}(m) =  \frac{V_D  V_P^{\rm NN}}{V_W^{\rm DD}} \!
\prod_{mv_j\notin\mathbb{Z}} \! 2 \inmod{\sin \pi mv_j} \,
\sum_{J=1}^{\tilde\chi(\theta^m_{\rm DD})} \inpar{{\rm Tr} \gamma_{m,p,J}}^2  \ .
\label{tadrrpp}
\eeq
To compute the RR tadpoles due to ${\ov{\D p}}$-branes
we just need to take into account the change of sign in the RR charge.
Concretely, the RR tadpoles from $\cc_{\bar p \bar p}$, and
$\cc_{p \bar p}$ plus $\cc_{\bar p p}$, are obtained from
(\ref{tadrrpp}) replacing $\inpar{{\rm Tr} \, \gamma_{m,p,J}}^2$ by
$\inpar{{\rm Tr} \, \gamma_{m,\bar p, J}}^2$ and
$-2{\rm Tr} \, \gamma_{m,p, J} {\rm Tr} \, \gamma_{m,\bar p, J}$ respectively.
Thus, the net effect of ${\ov{\D p}}$-branes amounts to substituting
\beq
\Tr \, \gamma_{m,p,J}  \to  \Tr \, \gamma_{m,p,J} - \Tr \, \gamma_{m,\bar p, J}
\label{ptobarp}
\eeq
in (\ref{tadrrpp}), as in orientifolds with supersymmetric $\Z_N$ action \cite{au}.

For our purposes, we do not need to analyze the generic $\cc_{pq}$, only
$\cc_{p9}$ in which NN or DN boundary conditions arise.
Coordinates $Y^i$ with mixed DN boundary conditions have oscillator expansions
with half-integer modes. For the fermions $\Psi^i$ world-sheet supersymmetry
then implies integer modes in NS and half-integer in R. In practice, the
upper characteristic of the $\vartheta$ functions in (\ref{tcylfer})
must be replaced by $\a + \oh$ for all DN directions. Effectively, $\cz_{p9}$
conforms to the amplitude of a sector twisted by $v_{p9}$ with entries
$\oh$ in the DN directions. Thus, necessarily the spin structure coefficients
must be that of a twisted sector. When $p=5$, we can take $v_{59}=\frac{N}2 v$,
so that $s_{0\oh}=-e^{-i\pi N S_v/2}$. On the other hand, for $p=7,3$,
$v_{p9}=\frac{N}4 v$ and  $s_{0\oh}=-e^{-i\pi N S_v/4}$.
These modified spin structure coefficients will show up in the RR tadpoles
because the $\a=0$, $\b=\oh$ term in the loop channel goes into RR in the
tree-level channel. Another relevant effect is the change of the GSO projection
for $p9$ states.

We now consider $p=5$ for concreteness.
If $Y^j$ is DN and $mv_j \in \Z$, the RR tadpole arising from $\cz_{59}(\theta^m)$ vanishes, 
the typical case being $m=0$. Otherwise we find: 
\beq 
T_{95+59}^{\rm RR}(m) =  2\,  e^{-i\pi \frac{N}2 S_v} V_D V_P^{\rm NN} \!
\prod_{{\rm NN} \atop {mv_j\notin \Z}} \! 2 \inmod{\sin{\inpar{\pi mv_j}}} \, 
\prod_{\rm DD} s(mv_j) \,
\sum_{J=1}^{\tilde\chi(\theta^m_{\rm DD})} \Tr \gamma_{m,9} \Tr \gamma_{m,5,J}  \ , 
\label{tadrr95} 
\eeq 
where we have defined 
\beq 
s(x)=\frac{\inmod{\sin \pi x}}{\sin \pi x} \quad ; \quad 
s(n) \equiv 0 \ . 
\label{signsin} 
\eeq
Notice that the DN (ND) directions for 59 (95) strings are the same as the DD directions 
for the corresponding 55 strings. All cylinder tadpoles have the same volume dependence 
if $mv_j \notin \Z$ for the DD coordinates, which in particular occurs when $m$ is odd 
and also when $m=\nm$. We can then sum the tadpoles to obtain 
\beq
\sum_{p,q=9,5} T_{pq}^{\rm RR}(m)  =  V_D V_P^{\rm NN}
\sqrt{ \frac{ \tilde\chi(\theta^m_{\rm NN})}{\tilde\chi(\theta^m_{\rm DD})}} \,  
\sum_{J=1}^{\tilde\chi(\theta^m_{\rm DD})} \left |\Tr \gamma_{m,9} +
\xi(mv) \, \sqrt{\tilde\chi(\theta^m_{\rm DD})} \Tr \gamma_{m,5,J} \right|^2 \ . 
\label{tadrr95sum} 
\eeq 
Here we have defined 
\beq 
\xi(mv)= e^{-i\pi \frac{N}2 S_v} \! \prod_{\rm DD} s(mv_j) \  .
\label{famsign} 
\eeq 
Also, by convention $\tilde\chi(\theta^m_{\rm NN}) \equiv 1$ when 
$mv_j \in \Z$ for the NN directions.

\subsection*{A.3 {\large{\bf M\"obius strip amplitudes}}}
%\subsection{M\"obius strip amplitudes}
\label{ssms}

The M\"obius strip amplitudes are given by:
\beq
\cam_p= \frac{V_D}{2N} \sum_{m=0}^N \int_0^\infty  \frac{dt}{t} \,
(8\pi^2 \a^\prime t)^{-D/2} \, \cz_p(\theta^m)  \  ,
\label{ampcyl}
\eeq
where
\beq
\cz_p(\theta^m) =  \Tr \inpar{P_{\rm GSO} \, \Omega_P
\, \theta^m \, e^{-2\pi t L_0} } \  .
\label{tracemobius}
\eeq
The trace is over open string states with $\D p$ branes at both endpoints.
Due to the $\Omega_P$ insertion the integral has divergences of type (\ref{treetad}) due
to massless states in the sector twisted by $\theta^{2m}$.
The resulting RR tadpoles will be denoted $T_p^{\rm RR}(\theta^m)$.

To evaluate $\cz_p(\theta^m)$ we start from $\cz_{pp}(\theta^m)$ and implement the action
of $\Omega_P$ inserted in the trace. Notice that being an open string amplitude,
$\Omega$ and $\op$ give the same result. For directions with NN boundary conditions
$\Omega$ acts on bosonic and fermionic oscillators, of frequency $r$,
as $\a_r \to e^{i\pi r} \a_r$ and $\psi_r \to e^{i\pi r} \psi_r$. In terms of the expansion
variable $q=e^{-2\pi t}$ this amounts to
\beq
q \to Q=e^{-i\pi} q \  .
\label{Qvar}
\eeq
Furthermore, $\Omega$ multiplies the NS vacuum by $e^{-i\pi/2}$ and the R vacuum by $-1$.
Then, leaving out possible sums over quantized momenta when $mv_j \in \Z$, the M\"obius
amplitude for $\D 9$-branes takes the form
\beq
\cz_9(\theta^m) = - \inpar{\Tr \gamma^{-1}_{\Omega m,9} \gamma^T_{\Omega m,9}}  \,
 \sum_{\a,\b=0,\oh} s_{\a\b}(0) \
\prod_{a=0}^3 \frac{-2 \sin \pi mv_a \eta_Q }{\vartheta_Q\left[\oh \atop {\oh + mv_a} \right] } \
\frac{\vartheta_Q\left[\a \atop {\b + mv_a} \right] }{\eta_Q} \ ,
\label{tmob9}
\eeq
where the subscript $Q$ means that the $\vartheta$ and $\eta$ functions
have $Q$ as variable. We can extract the divergence by performing modular 
transformations \cite{as}.
The resulting tadpole is given in (\ref{tadrrm9}).

For directions with DD boundary conditions there is an extra minus sign when
$\Omega$ acts on oscillators, i.e.
$\a_r \to -e^{i\pi r} \a_r$ and $\psi_r \to -e^{i\pi r} \psi_r$. This has the effect of
shifting by $\oh$ the lower characteristic of the $\vartheta$ functions involved
in the trace. To be more specific, consider the case of $\D 5$-branes. Then, as mentioned
previously, the vector $\frac{N}2 v$ has precisely half-integer entries in DD but
integer in the NN directions. We find that the consistent way of shifting the lower
characteristic is by adding  $\frac{N}2 v$. The trace will then vanish identically
in the supersymmetric cases as it should. Besides, changing the sign of $v$ gives
the same result as expected. In the end we arrive at the trace
\beqa
\cz_5(\theta^m) = &- & \sum_{J=1}^{\tilde\chi(\theta^m_{\rm DD})}
\inpar{\Tr \gamma^{-1}_{\Omega m,5,J}  \gamma^T_{\Omega m,5,J}}  \,
\nonumber \\[0.2cm]
& \times  &  \sum_{\a,\b=0,\oh} s_{\a\b}(0) \
\prod_{a=0}^3 \frac{-2 \sin \pi(m + \frac{N}2) v_a \eta_Q }
{\vartheta_Q\left[\oh \atop {\oh + mv_a + \frac{N}2 v_a} \right] }
\ \frac{\vartheta_Q\left[\a \atop {\b + mv_a} +  \frac{N}2 v_a \right] }{\eta_Q} \ .
\label{tmob5}
\eeqa
Here we have not written down possible sums over quantized momenta
when $mv_j \in \Z$ for NN directions
or over quantized windings when $mv_j \in \Z + \oh$ for DD directions.
The resulting tadpole turns out to be
\beqa
T_5^{\rm RR}(m)  & = &  -2^{\oh(D+I_P + I_W+2)} \,  \frac{V_D  V_P^{\rm NN}}{V_W^{\rm DD}}
\ \epsilon_5 \sum_{J=1}^{\tilde\chi(\theta^m_{\rm DD})} \Tr \gamma_{2m,5,J}
\nonumber \\[0.2cm]
& \times & \prod_{a=0}^3 c(mv_a + {\ts{\frac{N}2}} v_a) \,
\prod_{(m+ \frac{N}2)v_j\notin\Z} 2 \inmod{\sin{\inpar{\pi(m + \nm)v_j}}}
\label{tadrrm5}
\eeqa
where we have used (\ref{epsilondef}).

We finally work out some explicit results essential to compute the tadpole
cancellation condition when the $\Z_N$ action takes the form (\ref{v95N})
and there are O9 and O5-planes. Concretely, since we need the M\"obius tadpoles
due to massless states in a $\theta^{2\ell}$ sector, we will consider
$m=\ell$ and $m=\ell + \nm$ in the above expressions.
When $lv_j \notin \Z + \oh$, $\forall j$, all
M\"obius tadpoles $T_p^{\rm RR}(\ell+k)$, $k=0, \nm$, have the same
volume dependence $ V_D V_P^{\rm NN}$. Moreover,
they can be conveniently added according to the sector. Concretely,
\beqa
\sum_{p=9,5} T_p^{\rm RR}(\ell) & = & - 2^{\oh(D+I_P+2)} \, V_D
V_P^{\rm NN}\sqrt{ \frac{ \tilde\chi(\theta^\ell_{\rm NN})\,
\tilde\chi(\theta^{\ell+\frac{N}2}_{\rm
DD})}{\tilde\chi(\theta^{2\ell}_{\rm DD})}}\, \epsilon_9
\prod_{a=0}^3 c(\ell v_a)\, \times
\nonumber \\[0.2cm]
& & \sum_{J=1}^{\tilde\chi(\theta^\ell_{\rm DD})} \incor{\Tr
\gamma_{2\ell,9} - \epsilon_9 \epsilon_5  \, \xi(2\ell v) \,
\sqrt{\tilde\chi(\theta^{2\ell}_{\rm DD})} \Tr \gamma_{2\ell,5,J} } \ ;
\label{tadrr95mob1} \\[0.4cm]
\sum_{p=9,5} T_p^{\rm RR}(\ell+{\ts{\frac{N}2}}) & = & - 2^{\oh(D+I_P+2)} \, V_D
V_P^{\rm NN}\sqrt{ \frac{ \tilde\chi(\theta^\ell_{\rm NN})\,
\tilde\chi(\theta^{\ell}_{\rm
DD})}{\tilde\chi(\theta^{2\ell}_{\rm DD})}}\, \epsilon_9 \delta_9
\prod_{a=0}^3 c(\ell v_a + {\ts{\frac{N}2}} v_a)\, \times
\nonumber \\[0.2cm]
& & \sum_{J=1}^{\tilde\chi(\theta^{\ell+\frac{N}2}_{\rm DD})} 
\incor{\Tr \gamma_{2\ell,9} + \delta_9 \delta_5
\, \xi(2\ell v) \, \sqrt{\tilde\chi(\theta^{2\ell}_{\rm DD})} \Tr \gamma_{2\ell,5,J} } \ . 
\label{tadrr95mob2}
\eeqa 
In these expressions we have substituted the relation 
\beq 
\prod_{a=0}^3 c(\ell v_a + {\ts{\frac{N}2}} v_a)
\, c(\ell v_a) = -e^{-i\pi \frac{N}2 S_v} \! \prod_{\rm DD} s(2\ell v_j) = - \xi(2\ell v)\ . 
\label{csformula}
\eeq 
We have also used the trigonometric identity 
\beq \tilde\chi(\theta^{2\ell}_{\rm DD}) =
\tilde\chi(\theta^\ell_{\rm DD})\tilde\chi(\theta^{\ell+\frac{N}2}_{\rm DD}) 
\label{ctndformula} 
\eeq 
that
follows from the explicit form of $\theta$ and (\ref{chitil}). 
We further remark that for all crystallographic
actions of type (\ref{v95N}) and for $\ell$ such that $\ell v_1 \notin \Z + \oh$, there is an equality
$\tilde\chi(\theta^{2\ell}_{\rm NN}) = \tilde\chi(\theta^\ell_{\rm NN})$.

Clearly, the M\"obius tadpoles (\ref{tadrr95mob1}) and (\ref{tadrr95mob2})
are equally distributed over the fixed points of $\theta^{\ell}_{\rm DD}$
and $\theta^{\ell+\frac{N}2}_{\rm DD}$ respectively. The origin, labelled
by $J=1$, is a simultaneous fixed point so that its share of M\"obius
tadpoles, denoted $T_{\cam,1}^{\rm RR}(\ell)$, is a sum of contributions
from both sectors. We find
\beqa
T_{\cam,1}^{\rm RR}(\ell) \!\!\!\!
& = & \!\!\!\! - 2^{\oh(D+I_P+2)} \, V_D V_P^{\rm NN}
\sqrt{\frac{\tilde\chi(\theta^\ell_{\rm NN})}{\tilde\chi(\theta^{2\ell}_{\rm DD})}}\, \epsilon_9
\prod_{a=0}^3 c(\ell v_a)\,
\incor{\sqrt{\tilde\chi(\theta^{\ell+ \nm}_{\rm DD})} - \delta_9 \, \xi(2\ell v)
\sqrt{\tilde\chi(\theta^{\ell}_{\rm DD})} }
\nonumber \\[0.2cm]
& \times &   \incor{\Tr \gamma_{2\ell,9} + \xi(2\ell v) \,
\sqrt{\tilde\chi(\theta^{2\ell}_{\rm DD})} \Tr \gamma_{2\ell,5,1} } \ .
\label{tadrr95mob0}
\eeqa
Other common fixed points receive the same share.

\end{document}